\documentclass{egpubl}
\usepackage{eg2024}
 
\ConferencePaper

\CGFccby

\usepackage[T1]{fontenc}
\usepackage{dfadobe}  

\usepackage[export]{adjustbox}
\usepackage{graphicx}
\usepackage{color}
\usepackage{xcolor}
\usepackage{subcaption}
\usepackage{float}
\usepackage{tabu}
\usepackage{booktabs}
\usepackage{multirow}
\usepackage{amsmath}
\usepackage{listings}[language=C++]
\usepackage{hyperref}
\usepackage{overpic}
\usepackage[framemethod=tikz]{mdframed}
\usepackage{lipsum}
\usetikzlibrary{shadows}
\usepackage{egweblnk}
\usepackage{lineno}
\usepackage{tabularray}

\newmdenv[linewidth= 1pt,linecolor= white, tikzsetting={draw=black, line width = 2pt, dashed,%
dash pattern = on 4pt off 3pt}]{myshadowbox}
                
\usepackage{todonotes}

\usepackage{textcomp}
\usepackage{colortbl}
\usepackage{soul}

\soulregister\ref7
\soulregister\cite7

\definecolor{codegreen}{rgb}{0,0.6,0}
\definecolor{codegray}{rgb}{0.5,0.5,0.5}
\definecolor{codepurple}{rgb}{0.58,0,0.82}
\definecolor{backcolour}{rgb}{0.95,0.95,0.92}

\lstdefinestyle{mystyle}{ 
    commentstyle=\color{codegreen},
    keywordstyle=\color{magenta},
    numberstyle=\tiny\color{codegray},
    stringstyle=\color{codepurple},
    basicstyle=\ttfamily\small,
    breakatwhitespace=false,         
    breaklines=true,                 
    captionpos=b,                    
    keepspaces=true,                 
    numbers=left,                    
    showspaces=false,                
    showstringspaces=false,
    showtabs=false,                  
    tabsize=2
}
\title{SimLOD: Simultaneous LOD Generation and Rendering}

\author[M. Schütz \& L. Herzberger \& M. Wimmer]
{\parbox{\textwidth}{
    \centering Markus Schütz, Lukas Herzberger, Michael Wimmer
        }
        \\
{\parbox{\textwidth}{\centering TU Wien
       }
}
}

\biberVersion
\BibtexOrBiblatex
\usepackage[backend=biber,bibstyle=EG,citestyle=alphabetic,backref=true]{biblatex} 
\begin{document}

\teaser{
    \includegraphics[width=1.0\textwidth]{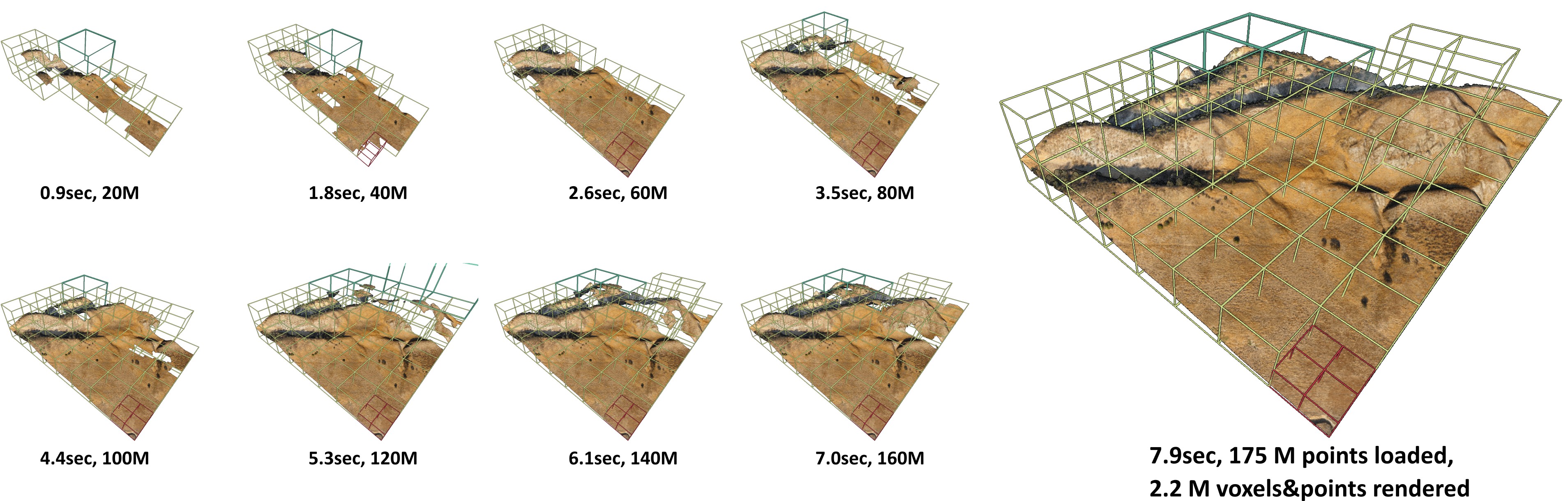}
    \centering
    \caption{State-of-the-art LOD generation approaches require users to wait until the entire data set is processed before they are able to view it. Our approach incrementally constructs an LOD structure directly on the GPU while points are being loaded from disk, and immediately displays intermediate results. Loading the depicted point cloud was bottlenecked to 22M points/sec by the industry-standard but CPU-intensive compression format (LAZ). Our approach is able to handle up to 580M points/sec while still rendering the loaded data in real time. }
    \label{fig:teaser}
}


\maketitle

\begin{abstract}

\textbf{About:}
We propose an incremental LOD generation approach for point clouds that allows us to simultaneously load points from disk, update an octree-based level-of-detail representation, and render the intermediate results in real time while additional points are still being loaded from disk. LOD construction and rendering are both implemented in CUDA and share the GPU's processing power, but each incremental update is lightweight enough to leave enough time to maintain real-time frame rates.

\vspace*{3px}
\noindent
\textbf{Background:} 
LOD construction is typically implemented as a preprocessing step that requires users to wait before they are able to view the results in real time. This approach allows users to view intermediate results right away.

\vspace*{3px}
\noindent
\textbf{Results:} 
Our approach is able to stream points from an SSD and update the octree on the GPU at rates of up to 580 million points per second (\textasciitilde9.3GB/s from a PCIe 5.0 SSD) on an RTX 4090. Depending on the data set, our approach spends an average of about 1 to 2 ms to incrementally insert 1 million points into the octree, allowing us to insert several million points per frame into the LOD structure and render the intermediate results within the same frame.

\vspace*{3px}
\noindent
\textbf{Discussion/Limitations:} 
We aim to provide near-instant, real-time visualization of large data sets without preprocessing. Out-of-core processing of arbitrarily large data sets and color-filtering for higher-quality LODs are subject to future work.



\end{abstract}


\clearpage

\section{Introduction}

Point clouds are an alternative representation of 3D models, comprising vertex-colored points without connectivity, and are typically obtained by scanning the real world via means such as laser scanners or photogrammetry. Since they are vertex-colored, large amounts of points are required to represent details that triangle meshes can cheaply simulate with textures. As such, point clouds are not an efficient representation for games, but they are nevertheless popular and ubiquitously available due to the need to scan real-world objects, buildings, and even whole countries. 

Examples for massive point-cloud data sets include: The 3D Elevation Program (3DEP), which intends to scan the entire USA \cite{USGS:3DEP}, and Entwine\cite{Entwine}, which currently hosts 53.6 trillion points that were collected in various individual scan campaigns within the 3DEP program~\cite{USGS:Entwine}. The Actueel Hoogtebestand Nederland (AHN)~\cite{AHN} program repeatedly scans the entire Netherlands, with the second campaign resulting in 640 billion points~\cite{AHN2}, and the fourth campaign being underway. Many other countries also run their own country-wide scanning programs to capture the current state of land and infrastructure. At a smaller scale, buildings are often scanned as part of construction, planning, and digital heritage. But even though these are smaller in extent, they still comprise hundreds of millions to several billion points due to the higher scan density of terrestrial LIDAR and photogrammetry. 

One of the main issues when working with large point clouds is the computational effort that is required to process and render hundreds of millions to billions of points. Level-of-detail structures are an essential tool to quickly display visible parts of a scene up to a certain amount of detail, thus reducing load times and improving rendering performance on lower-end devices. However, generating these structures can also be a time-consuming process. Recent GPU-based methods~\cite{SCHUETZ-2023-LOD} improved LOD compute times down to a second per billion points, but they still require users to wait until the entire data set has been loaded and processed before the resulting LOD structure can be rendered. Thus, if loading a billion points takes 60 seconds plus 1 second of processing, users still have to wait 61 seconds to inspect the results.

In this paper, we propose an incremental LOD generation approach that allows users to instantly look at data sets as they are streamed from disk, without the need to wait until LOD structures are generated in advance. This approach is currently in-core, i.e., data sets must fit into memory, but we expect that it will serve as a basis for future out-of-core implementations to support arbitrarily large data sets. 

Our contributions to the state-of-the-art are as follows:

\begin{itemize}
    \item An approach that instantly displays large amounts of points as they are loaded from fast SSDs, and simultaneously updates an LOD structure directly on the GPU to guarantee high real-time rendering performance.
    \item As a smaller, additional contribution, we demonstrate that dynamically growing arrays of points via linked-lists of chunks can be rendered fairly efficiently in modern, compute-based rendering pipelines.
\end{itemize}

Specifically not a contribution is the development of a new LOD structure. We generate the same structure as Wand et al.~\cite{Wand2008} or Schütz et al.~\cite{SCHUETZ-2023-LOD}, which are also very similar to the widely used modifiable nested octrees~\cite{scheiblauer2011}. We opted for constructing the former over the latter because quantized voxels compress better than full-precision points (down to 10 bits per colored voxel), which improves the transfer speed of lower LODs over the network. Furthermore, since inner nodes are redundant, we can compute more representative, color-filtered values. However, both compression and color filtering are applied in post-processing before storing the results on disk and are not covered by this paper. This paper focuses on incrementally creating the LOD structure and its geometry as fast as possible for immediate display and picks a single color value from the first point that falls into a voxel cell.

\section{Related Work}

\subsection{LOD Structures for Point Clouds}

Point-based and hybrid LOD representations were initially proposed as a means to efficiently render mesh models at lower resolutions~\cite{QSplat, 964491, 10.2312:EGWR:EGWR02:043-052, Dachsbacher2003} and possibly switch to the original triangle model at close-up views. With the rising popularity of 3D scanners that produce point clouds as intermediate and/or final results, these algorithms also became useful to handle the enormous amounts of geometry that are generated by scanning the real world. Layered point clouds (LPC)~\cite{GOBBETTI2004} was the first GPU-friendly as well as view-dependent approach, which made it suitable for visualizing arbitrarily large data sets. LPCs organize points into a multi-resolution binary tree where each node represents a part of the point cloud at a certain level of detail, with the root node depicting the whole data set at a coarse resolution, and child nodes adding additional detail in their respective regions. Since then, further research has improved various aspects of LPCs, such as utilizing different tree structures~\cite{InstantPoints, Wand2008, Goswami2010, NestedIndexing}, improving LOD construction times~\cite{MartinezRubi2015, Kang2019, SCHUETZ-2020-MPC, Bormann:PCI, 9671659} and higher-quality sampling strategies instead of selecting random subsets~\cite{VANOOSTEROM2022119, SCHUETZ-2023-LOD}.

\begin{figure}
    \centering
    \begin{subfigure}[t]{0.49\columnwidth}
        \includegraphics[width=\columnwidth]{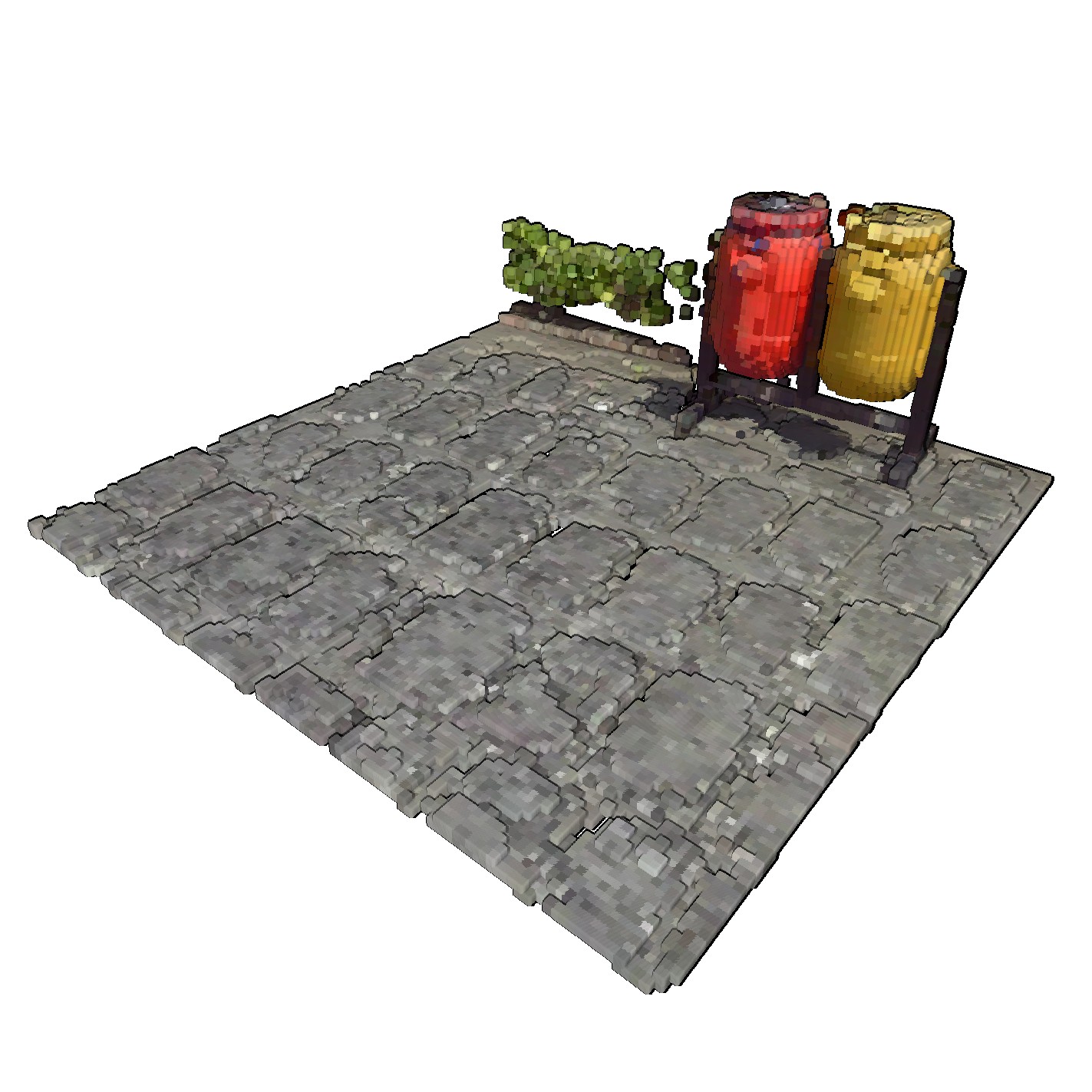}
        \caption{Inner node with Voxels.}
    \end{subfigure}
    \hfill
    \begin{subfigure}[t]{0.49\columnwidth}
        \includegraphics[width=\columnwidth]{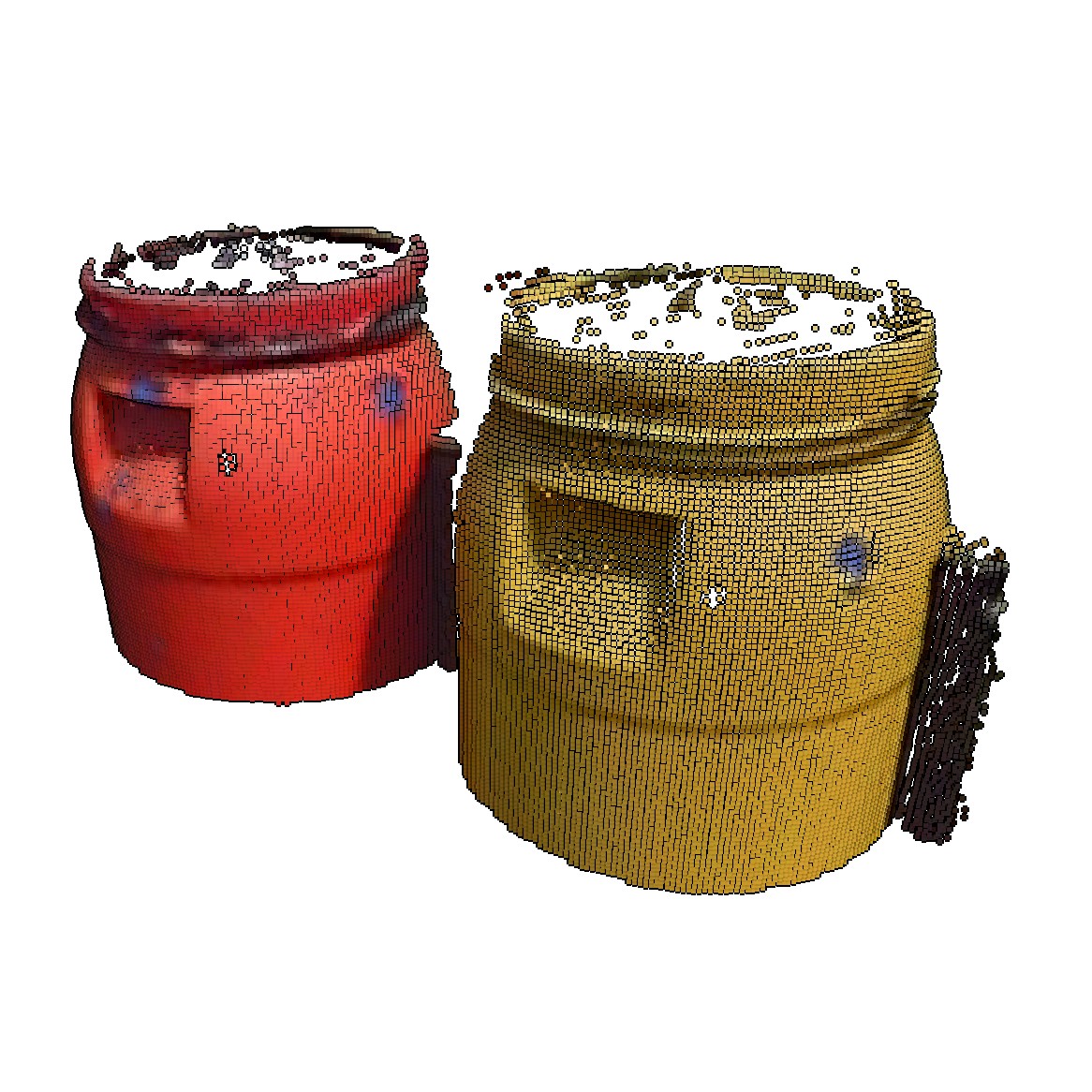}
        \caption{Leaf node with points.}
    \end{subfigure}
    
    \caption{(a) Close-up of a lower-resolution inner-node comprising 20\,698 voxels that were sampled on a $128^3$ grid. (b) A full-resolution leaf node comprising 22\,858 points.}
    \label{fig:lod_inner_leaf}
\end{figure}

In this paper, we focus on constructing a variation of LPCs proposed by Wand et al.~\cite{Wand2008}, which utilizes an octree where each node creates a coarse representation of the point cloud with a resolution of $128^3$ cells, and leaf nodes store the original, full-precision point data, as shown in Figure~\ref{fig:lod_inner_leaf}. Wand et al. suggest various primitives as coarse, representative samples (quantized points, Surfels, ...), but for this work we consider each cell of the $128^3$ grid to be a voxel. A similar voxel-based LOD structure by Chajdas et al.~\cite{Chajdas2014Scalable} uses $256^3$ voxel grids in inner nodes and original triangle data in leaf nodes. Modifiable nested octrees (MNOs)~\cite{scheiblauer2011} are also similar to the approach by Wand et al.~\cite{Wand2008}, but instead of storing all points in leaves and representative samples (Surfels, Voxels, ...) in inner nodes, MNOs fill empty grid cells with points from the original data set. 

Since our goal is to display all points the instant they are loaded from disk to GPU memory, we need LOD construction approaches that are capable of efficiently inserting new points into the hierarchy, expanding it if necessary, and updating all affected levels of detail. This disqualifies recent bottom-up or hybrid bottom-up and top-down approaches~\cite{MartinezRubi2015, SCHUETZ-2020-MPC, Bormann:PCI, SCHUETZ-2023-LOD} that achieve a high construction performance, but which require preprocessing steps that iterate through all data before they actually start with the construction of the hierarchy. Wand et al.~\cite{Wand2008} as well as Scheiblauer and Wimmer~\cite{scheiblauer2011}, on the other hand, propose modifiable LOD structures with deletion and insertion methods, which make these inherently suitable to our goal since we can add a batch of points, draw the results, and then add another batch of points. Bormann et al.~\cite{RealTimeIndexingBormann} were the first to specifically explore this concept for point clouds by utilizing MNOs, but flushing updated octree nodes to disk that an external rendering engine can then stream and display. They achieved a throughput of 1.8 million points per second, which is sufficient to construct an LOD structure as fast as a laser scanner generates point data. A downside of these CPU-based approaches is that they do not parallelize well, as threads need to avoid processing the same node or otherwise sync critical operations. In this paper, we propose a GPU-friendly approach that allows an arbitrary amount of threads to simultaneously insert points, which allows us to construct and render on the same GPU at rates of up to 580 million points per second, or up to 1.2 billion points per second for construction without rendering. 

While we focus on point clouds, there are some notable related works in other fields that allow simultaneous LOD generation and rendering. In general, any LOD structure with insertion operations can be assumed to fit these criteria, as long as inserting a meaningful amount of geometry can be done in milliseconds. Careil et al.~\cite{ModifyVoxels} demonstrate a voxel painter that is backed by a compressed LOD structure. We believe that \emph{Dreams} -- a popular 3D scene painting and game development tool for the PS4 -- also matches the criteria, as developers reported experiments with LOD structures, and described the current engine as a ``cloud of clouds of point clouds"~\cite{2015learning}. 

\subsection{Linked Lists}

Linked lists are a well-known and simple structure whose constant insertion and deletion complexity, as well as the possibility to dynamically grow without relocation of existing data, make it useful as part of more complex data structures and algorithms (e.g.least-recently-used (LRU) Caches~\cite{BufferCacheManagement}). On GPUs, they can be used to realize order-independent transparency~\cite{yang_real-time_2010} by creating pixel-wise lists of fragments that can then be sorted and drawn front to back. In this paper, we use linked lists to append an unknown amount of points and voxels to octree nodes during LOD construction. 







\section{Data Structure}

\subsection{Octree}

\begin{figure}
    \centering
    \includegraphics[width=\columnwidth]{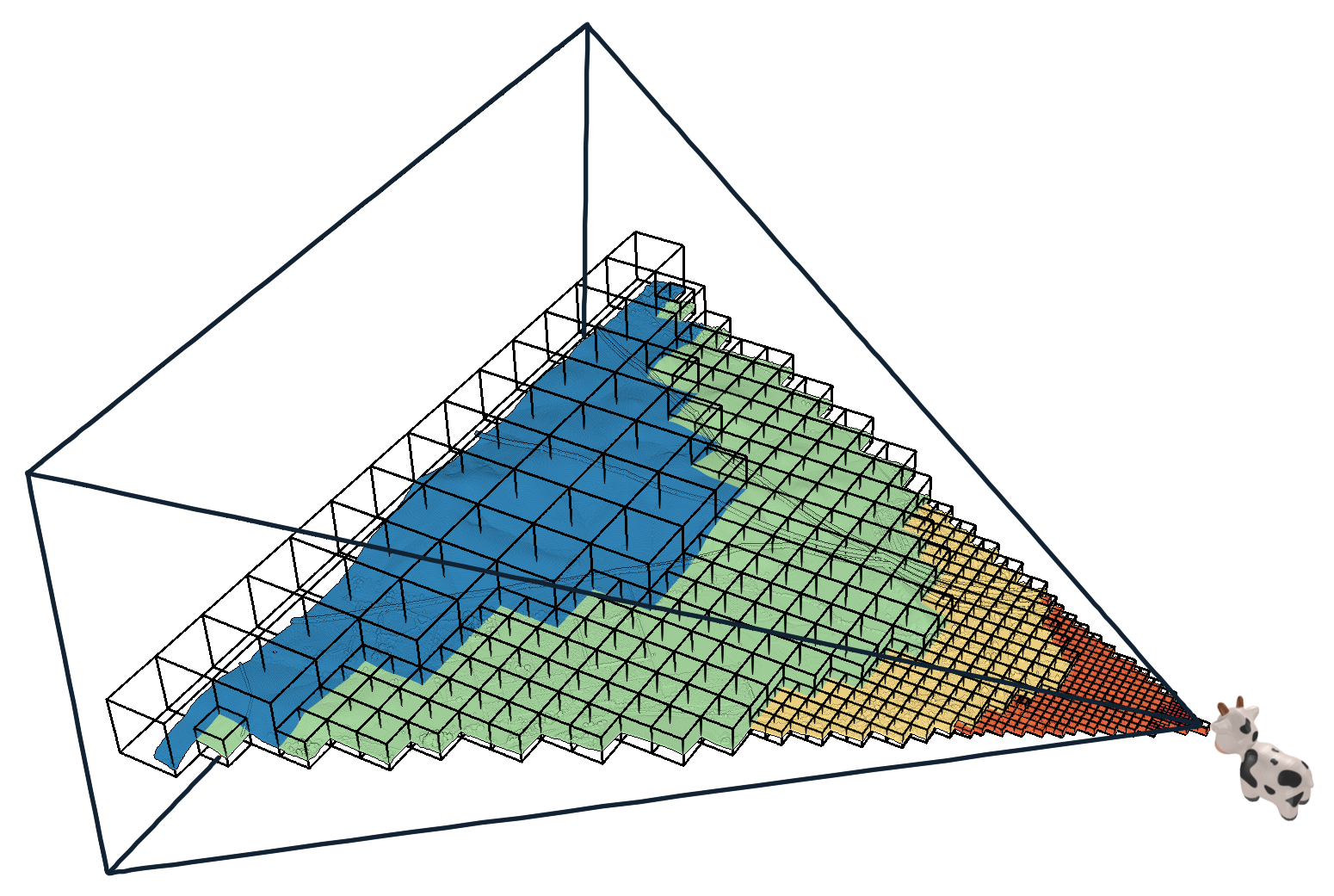}
    \caption{Higher-level octree nodes rendered closer to the camera. Inner nodes store lists of representative voxels that were sampled on a $128^3$ grid, and leaf nodes store up to 50k full-precision points.}
    \label{fig:lod_frustum}
\end{figure}

The LOD data structure we use is an octree-based~\cite{scheiblauer2011} layered point cloud~\cite{GOBBETTI2004} with representative voxels in inner nodes and the original, full-precision point data in leaf nodes, which makes it essentially identical to the structures of Wand et al.~\cite{Wand2008} or Schütz et al.~\cite{SCHUETZ-2023-LOD}. Leaf nodes store up to 50k points and inner nodes up to $128^3$ (2M) voxels, but typically closer to $128^2$ (16k) voxels due to the surfacic nature of point cloud data sets. The sparse nature of surface voxels is the reason why we store them in lists instead of grids -- exactly the same as points. Figure~\ref{fig:lod_frustum} illustrates how more detailed, higher-level nodes are rendered close to the camera. 

\begin{figure}
    \centering
    \includegraphics[width=\columnwidth]{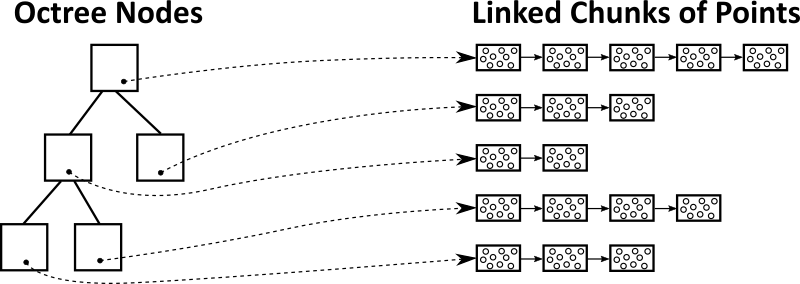}
    \caption{Octree nodes store 3D data as linked chunks of points/voxels. Linked chunks enable dynamic growth as new 3D data is added; efficient removal (after splitting leaves) by simply putting chunks back into a pool for re-use; and efficient rendering in compute-based pipelines.}
    \label{fig:nodes_n_chunks}
\end{figure}

The difference to the structure of Schütz et al.~\cite{SCHUETZ-2023-LOD} is that we store points and voxels in linked lists of chunks of points, which allows us to add additional capacity by allocating and linking additional chunks, as shown in Figure~\ref{fig:nodes_n_chunks}. An additional difference to Wand et al.~\cite{Wand2008} is that they use hash maps for their $128^3$ voxel sampling grids, whereas we use a $128^3 bit = 256kb$ occupancy grid per inner node to simplify massivelly parallel sampling on the GPU. 

Despite the support for dynamic growth via linked lists, this structure still supports efficient rendering in compute-based pipelines, where each individual workgroup can process points in a chunk in parallel, and then traverse to the next chunk as needed. In our implementation, each chunk stores up to $1,000$ points or voxels (Discussion in Section~\ref{sec:chunkSizes}), with the latter being implemented as points where coordinates are quantized to the center of a voxel cell.

\subsection{Persistent Buffer}

Since we require large amounts of memory allocations on the device from within the CUDA kernel throughout the LOD construction over hundreds of frames, we manage our own custom persistent buffer to keep the cost of memory allocation to a minimum. To that end, we simply pre-allocate 90\% of the available GPU memory. An atomic offset counter keeps track of how much memory we already allocated and is used to compute the returned memory pointer during new allocations. 

Note that sparse buffers via virtual memory management may be an alternative, as discussed in Section~\ref{sec:conclusion}.

\subsection{Voxel Sampling Grid}

Voxels are sampled by inscribing a $128^3$ voxel grid into each inner node, using 1 bit per cell to indicate whether that cell is still empty or already occupied. Inner nodes therefore require 256kb of memory in addition to the chunks storing the voxels (in our implementation as points with quantized coordinates). Grids are allocated from the persistent buffer whenever a leaf node is converted into an inner node. 

\subsection{Chunks and the Chunk Pool}

We use chunks of points/voxels to dynamically increase the capacity of each node as needed, and a chunk pool where we return chunks that are freed after splitting a leaf node (chunk allocations for the newly created inner node are handled separately). Each chunk has a static capacity of \emph{N} points/voxels ($1,000$ in our implementation), which makes it trivial to manage chunks as they all have the same size. Initially, the pool is empty and new chunks are allocated from the persistent buffer. When chunks are freed after splitting a leaf node, we store the pointers to these chunks inside the chunk pool. Future chunk allocations first attempt to acquire chunk pointers from the pool, and only allocate new chunks from the persistent buffer if there are none left in the pool.



\section{Incremental LOD Construction -- Overview}

\begin{figure*}
    \centering
    \includegraphics[width=\textwidth]{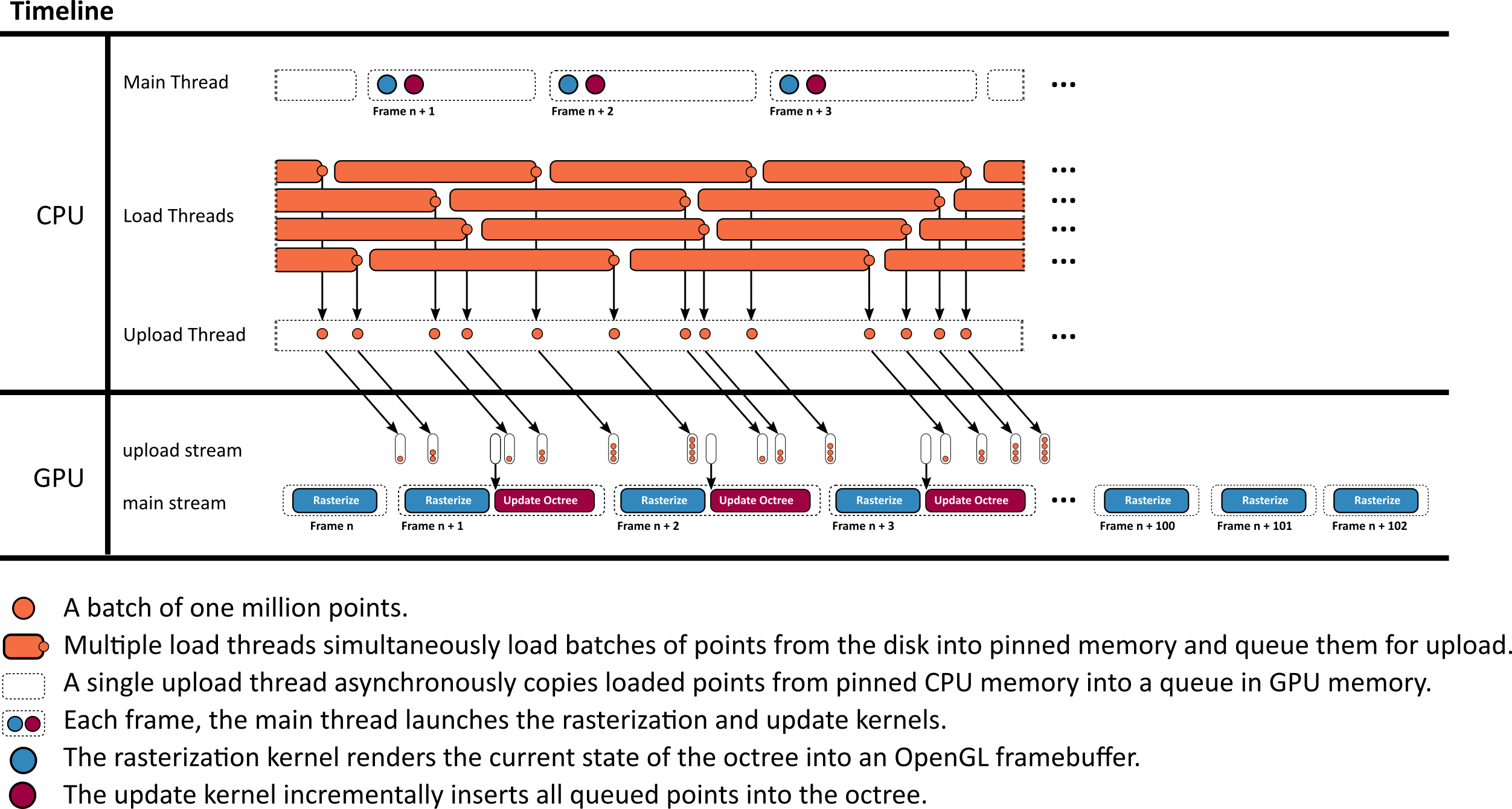}
    \caption{Timeline of our system over several frames. }
    \label{fig:timeline}
\end{figure*}

Our method loads batches of points from disk to GPU memory, updates the LOD structure in one CUDA kernel, and renders the updated results with another CUDA kernel. Figure~\ref{fig:timeline} shows an overview of that pipeline. Both kernels utilize persistent threads~\cite{PersistentThreads, GpuSoftwarePipeline} using the cooperative group API~\cite{CooperativeGroups} in order to merge numerous sub-passes into a single CUDA kernel. Points are loaded from disk to pinned CPU memory in batches of 1M points, utilizing multiple load threads. Whenever a batch is loaded, it is appended to a queue. A single uploader thread watches that queue and asynchronously copies any loaded batches to a queue in GPU memory. In each frame, the main thread launches the rasterize kernel that draws the entire scene, followed by an update kernel that incrementally inserts all batches of points into the octree that finished uploading to the GPU. 

\section{Incrementally Updating the Octree}

In each frame, the GPU may receive several batches of 1M points each. The update kernel loops through the batches and inserts them into the octree as shown in Figure~\ref{fig:kernel_update}. First, the octree is expanded until the resulting leaf nodes will hold at most 50k points. It then traverses each point of the batch through the octree again to generate voxels for inner nodes. Afterwards, it allocates sufficient chunks for each node to store all points in leaf-, and voxels in inner nodes. In the last step, it inserts the points and voxels into the newly allocated chunks of memory.

The premise of this approach is that it is cheaper in massively parallel settings to traverse the octree multiple times for each point and only insert them once at the end, rather than traversing the tree once per point but with the need for complex synchronization mechanisms whenever a node needs splitting or additional chunks of memory need to be allocated. 

\subsection{Expanding the Octree}

CPU-based top-down approaches~\cite{Wand2008, scheiblauer2011} typically traverse the hierarchy from root to leaf, update visited nodes along the way, and append points to leaf nodes. If a leaf node receives too many points, it ``spills" and is split into 8 child nodes. The points inside the spilling node are then redistributed to its newly generated child nodes. This approach works well on CPUs, where we can limit the insertion and expansion of a subtree to a single thread, but it raises issues in a massively parallel setting, where thousands of threads may want to insert points while we simultaneously need to split that node and redistribute the points it already contains. 

To support massively parallel insertions of all points on the GPU, we propose an iterative approach that resembles a depth-first-iterative-deepening search~\cite{korf1985depth}. Instead of attempting to fully expand the octree structure in a single step, we repeatedly expand it by one level until no more expansions are needed. This approach also decouples expansions of the hierarchy and insertions into a node's list of points, which is now deferred to a separate pass. Since we already defer the insertion of points into nodes, we also defer the redistribution of points from spilled nodes. We maintain a spill buffer, which accumulates points of spilled nodes. Points in the spill buffer are subsequently treated exactly the same as points inside the batch buffer that we are currently adding to the octree, i.e., the update kernel reinserts spilled points into the octree from scratch, along with the newly loaded batch of points. 

\begin{figure*}[h!]
    \centering
    \begin{subfigure}[t]{\textwidth}
        \includegraphics[width=\textwidth]{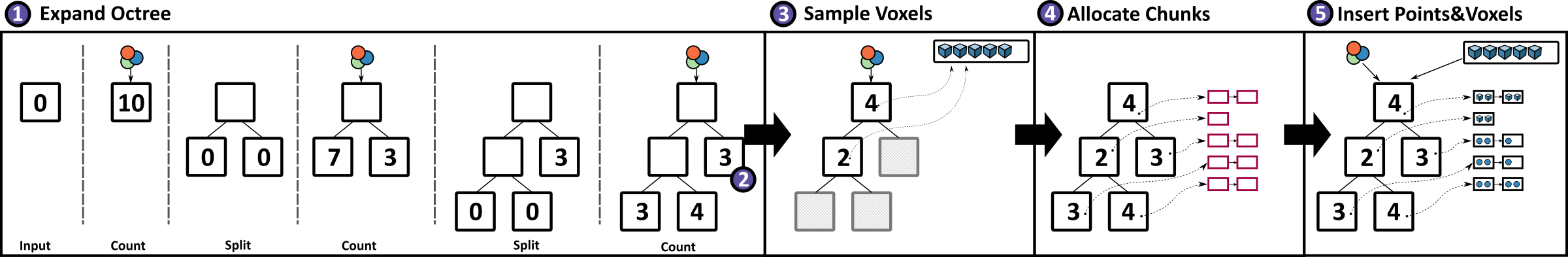}
        \caption{Adding 10 points to the octree. (1) Expanding the octree by repeatedly counting and splitting until leaf nodes hold at most \emph{T} points (depicted: 5, in practice: 50k).
(2) Leaves that were not split do not count points again. 
(3) The voxel sampling pass inserts all points again, creates voxels for empty cells in inner nodes, and stores new voxels (and the nodes they belong to) in a temporary \emph{backlog} buffer.
(4) Now that we know the number of new points and voxels, we allocate the necessary chunks (depicted size: 2, in practice: 1000) to store them.
(5) All points are inserted again, traverse to the leaf, and are inserted into the chunks. Voxels from the \emph{backlog} are inserted into the respective inner nodes. }
    \end{subfigure}
    \begin{subfigure}[t]{\textwidth}
        \includegraphics[width=\textwidth]{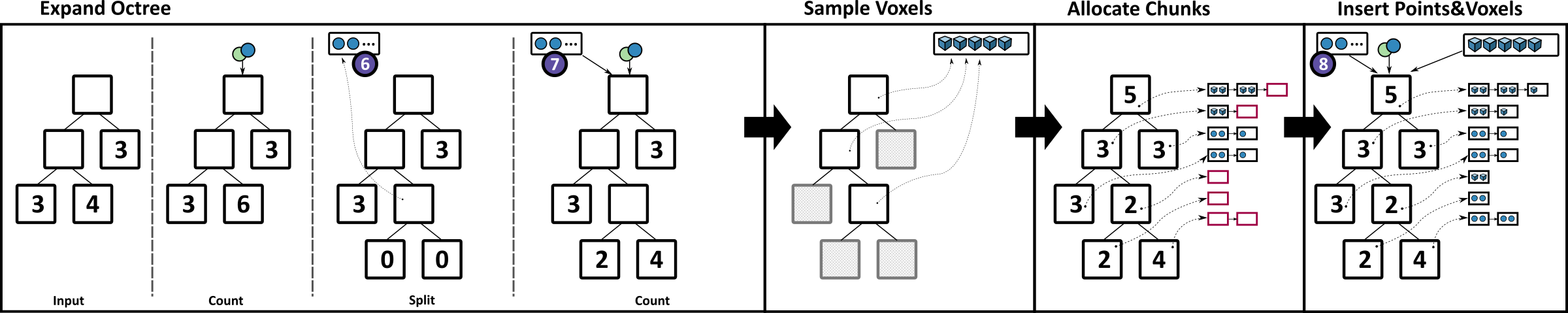}
        \caption{For illustrative purposes, we now add a batch of just two points which makes one of the nodes spill. (6) When splitting, we move all previously inserted points into a spill buffer. (7, 8) for the remainder of this frame's update, points in the spill buffer and the current batch get identical treatment. }
    \end{subfigure}
    \caption{The CUDA kernel that incrementally updates the octree. (a) First, it inserts a batch with 10 points into the initially empty octree and (b) then adds another batch with two points that causes a split of a non-empty leaf node.}
    \label{fig:kernel_update}
\end{figure*}

In detail, to expand the octree, we repeat the following two sub-passes until no more nodes are spilled and all leaf nodes are marked as final for this update (see also Figure~\ref{fig:kernel_update}): 

\begin{itemize}
    \item \textbf{Counting:} In each iteration, we traverse the octree for each point of the batch and all spilled points accumulated in previous iterations during the current update, and atomically increment the point counter of each hit leaf node that is not yet marked as final by one.
    \item \textbf{Splitting:} All leaf nodes whose point counter exceeds a given threshold, e.g., 50k points, are split into 8 child nodes, each with a point counter of $0$. The points it already contained are added to the list of spilled points. Note that the spilled points do not need to be associated with the nodes that they formerly belonged to -- they are added to the octree from scratch. Furthermore, the chunks that stored the spilled points are released back to the chunk pool and may be acquired again later. Leaf nodes whose point counter does not exceed the threshold are marked as final so that further iterations during this update do not count points twice.
\end{itemize}

The expansion pass is finished when no more nodes are spilling, i.e., all leaf nodes are marked final. 

\subsection{Voxel Sampling}

Lower levels of detail are populated with voxel representations of the points that traversed these nodes. Therefore, once the octree expansion is finished, we traverse each point through the octree again, and whenever a point visits an inner node, we project it into the inscribed $128^3$ voxel sampling grid and check if the respective cell is empty or already occupied by a voxel. If the cell is empty, we create a voxel, increment the node's voxel counter, and set the corresponding bit in the sample grid to mark it as occupied. Note that in this way, the voxel gets the color of the first point that projects to it. 

However, just like the points, we do not store voxels in the nodes right away because we do not know the amount of memory/chunks that each node requires until all voxels for the current incremental update are generated. Thus, voxels are first stored in a temporary backlog buffer with a large capacity. In theory, adding a batch of 1 million points may produce up to $(octreeLevels - 1)$ million voxels because each inner node's sampling grid has the potential to hold $128^3 = 2M$ voxels, and adding spatially close points may lead to several new octree levels until they are all separated into leaf nodes with at most 50k points. However, in practice, none of the test data sets of this paper produced more than 1M voxels per batch of 1M points, and of our numerous other data sets, the largest required backlog size was 2.4M voxels. Thus, we suggest using a backlog size of 10M points to be safe.

\subsection{Allocating Chunks}

After expansion and voxel sampling, we now know the exact amount of points and voxels that we need to store in leaf and inner nodes. Using this knowledge, we check all affected nodes whether their chunks have sufficient free space to store the new points/voxels, or if we need to allocate new chunks of memory to raise the nodes' capacity by 1000 points or voxels per chunk. In total, we need $\lfloor \frac{counter + POINTS\_PER\_CHUNK - 1}{POINTS\_PER\_CHUNK} \rfloor$ linked chunks per node. 

\subsection{Storing Points and Voxels}

To store points inside nodes, we traverse each point from the input batch and the spill buffer again through the octree to the respective leaf node and atomically update that node's \emph{numPoints} variable. The atomic update returns the point index within the node, from which we can compute the index of the chunk and the index within the chunk where we store the point.

We then iterate through the voxels in the backlog buffer, which stores voxels and for each voxel a pointer to the inner node that it belongs to. Insertion is handled the same way as points -- we atomically update each node's \emph{numVoxels} variable, which returns an index from which we can compute the target chunk index and the position within that chunk.

\section{Rendering}

Points and voxels are both drawn as pixel-sized splats by a CUDA kernel that utilizes atomic operations to retain the closest sample in each pixel~\cite{Gnther2013AGP, 2015learning, SCHUETZ-2022-PCC}. Custom compute-based software-rasterization pipelines are particularly useful for our method because traditional vertex-shader-based pipelines are not suitable for drawing linked lists of chunks of points. A CUDA kernel, however, has no issues looping through points in a chunk, and then traversing to the next chunk in the list. The recently introduced mesh and task shaders could theoretically also deal with linked lists of chunks of points, but they may benefit from smaller chunk sizes, and perhaps even finer-grained nodes (smaller sampling grids that lead to fewer voxels per node, and a lower maximum of points in leaf nodes).

During rendering, we first assemble a list of visible nodes, comprising all nodes whose bounding box intersects the view frustum and which have a certain size on screen. Since inner nodes have a voxel resolution of $128^3$, we need to draw their half-sized children if they grow larger than 128 pixels. We draw nodes that fulfill the following conditions:

\begin{itemize}
    \item Nodes that intersect the view frustum.
    \item Nodes whose parents are larger than 128 pixels. In that case, the parent is hidden and all its children are made visible instead. 
\end{itemize}

Figure~\ref{fig:lod_frustum} illustrates the resulting selection of rendered octree nodes within a frustum. Seemingly higher-LOD nodes are rendered towards the edge of the screen due to perspective distortions that make the screen-space bounding boxes bigger. For performance-sensitive applications, developers may instead want to do the opposite and reduce the LOD at the periphery and fill the resulting holes by increasing the point sizes. 

To draw points or voxels, we launch one workgroup per visible node whose threads loop through all samples of the node and jump to the next chunk when needed, as shown in listing~\ref{lst:render_chunks}.

\begin{lstlisting}[language=Java,label={lst:render_chunks},caption={CUDA code showing threads of a workgroup iterating through all points in a node, processing \emph{num\_threads} points at a time in parallel. Threads advance to the next chunk as needed.},captionpos=b]
Node* node = &visibleNodes[workgroupIndex];
Chunk* chunk = node->points;
int chunkIndex = 0;

for(
    int pointIndex = block.thread_rank(); 
    pointIndex < node->numPoints; 
    pointIndex += block.num_threads()
){
    int targetChunkIndex = pointIndex / POINTS_PER_CHUNK;
    
    if(chunkIndex < targetChunkIndex){
        chunk = chunk->next;
        chunkIndex++;
    }
    
    int pIndex = pointIndex % POINTS_PER_CHUNK;
    Point point = chunk->points[pIndex];
    
    rasterize(point);
}
\end{lstlisting}

\section{Evaluation}
\label{sec:evaluation}

Our method was implemented in C++ and CUDA, and evaluated on the test data sets shown in Figure~\ref{fig:test_data_sets}.

\begingroup
\setlength{\tabcolsep}{1pt}
\renewcommand{\arraystretch}{0.1}
\begin{figure*}
\begin{tabular*}{\textwidth}{m{1em}cccc}
 & Overview (color) & Overview (nodes) & Closeup (color) & Closeup (nodes) \vspace*{0.5mm} \\ 
 \rotatebox{90}{Chiller}  &
     \raisebox{-0.5\totalheight}{\includegraphics[width=0.24\textwidth]{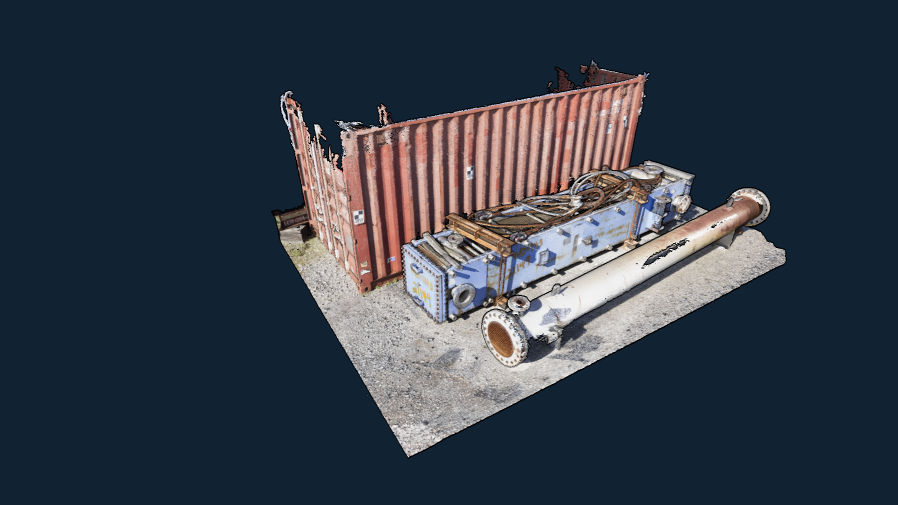}} & 
     \raisebox{-0.5\totalheight}{\includegraphics[width=0.24\textwidth]{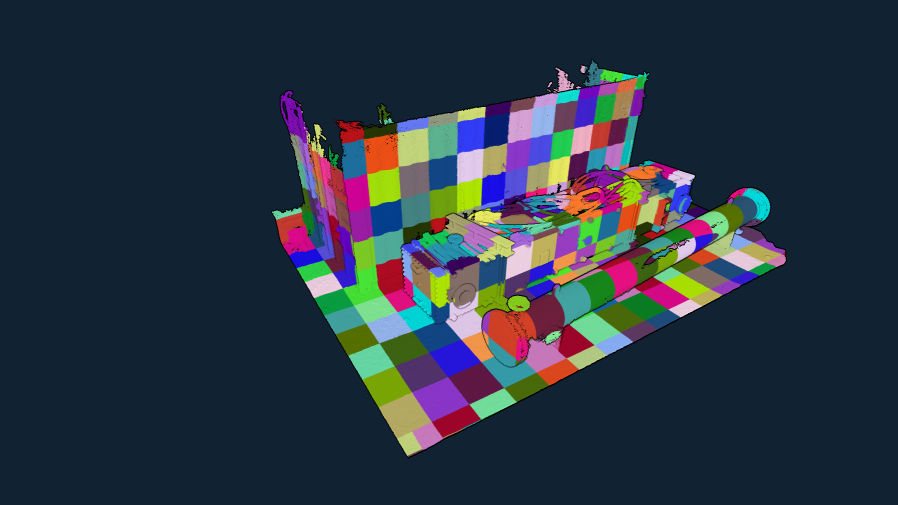}} &
     \raisebox{-0.5\totalheight}{\includegraphics[width=0.24\textwidth]{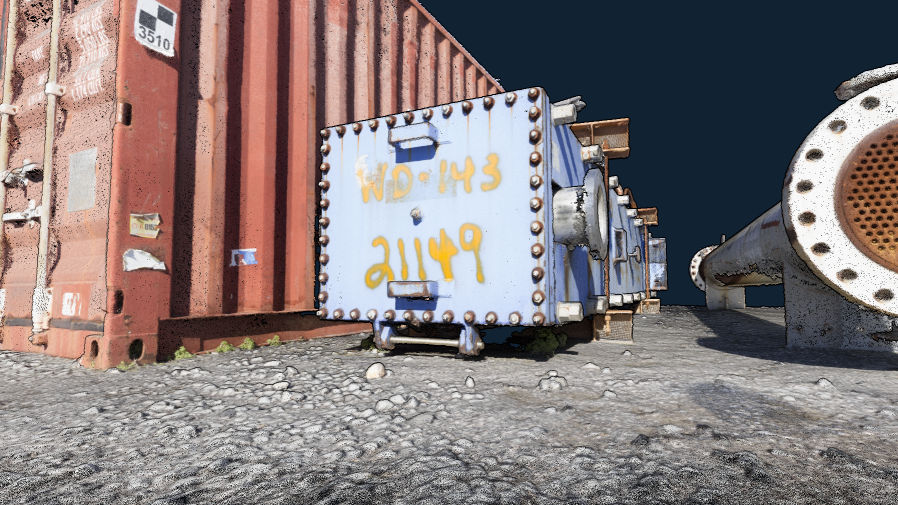}} &
     \raisebox{-0.5\totalheight}{\includegraphics[width=0.24\textwidth]{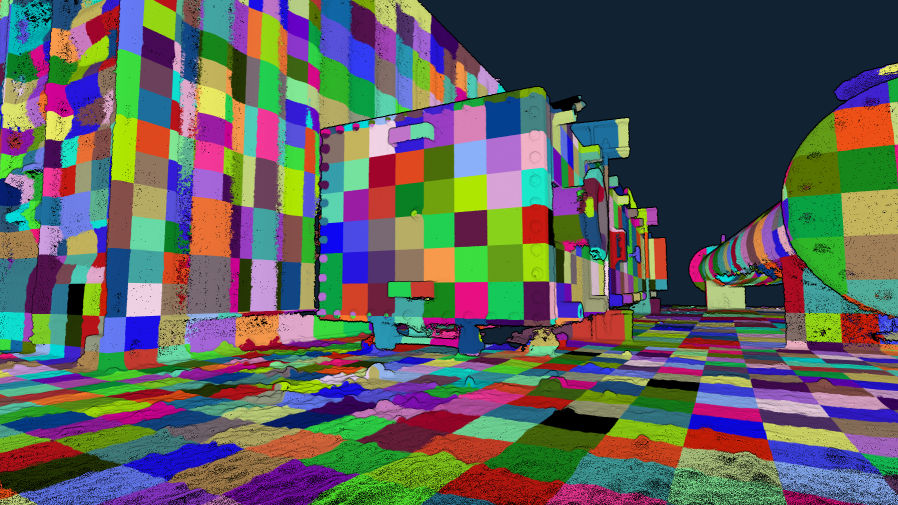}} \vspace*{0.5mm}
 \\ 
 \rotatebox{90}{Retz}  &
     \raisebox{-0.5\totalheight}{\includegraphics[width=0.24\textwidth]{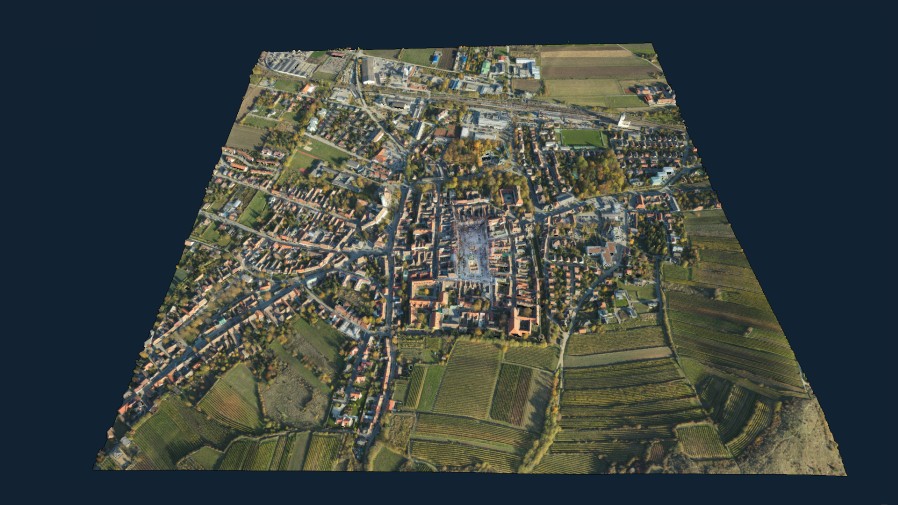}} & 
     \raisebox{-0.5\totalheight}{\includegraphics[width=0.24\textwidth]{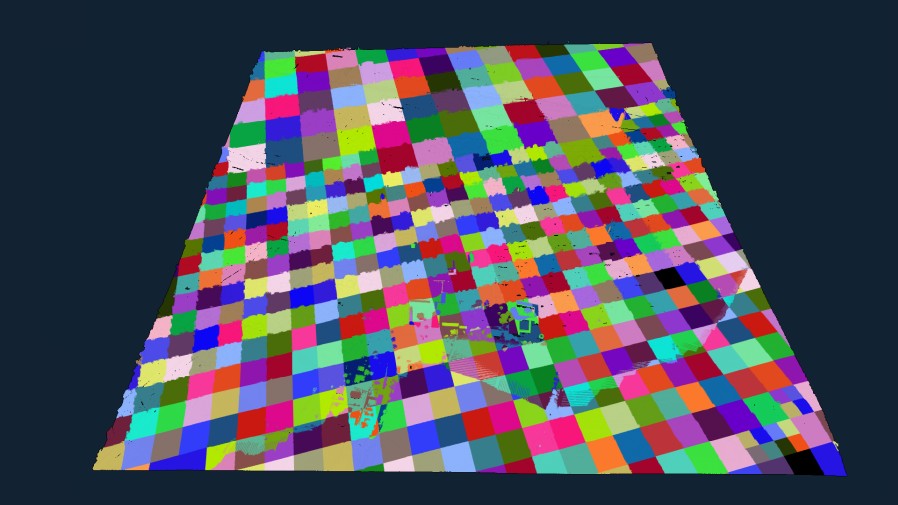}} &
     \raisebox{-0.5\totalheight}{\includegraphics[width=0.24\textwidth]{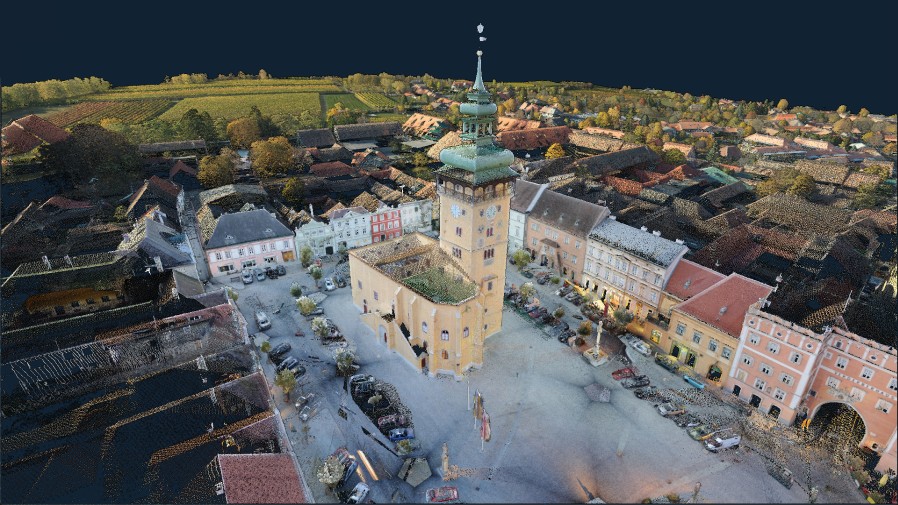}} &
     \raisebox{-0.5\totalheight}{\includegraphics[width=0.24\textwidth]{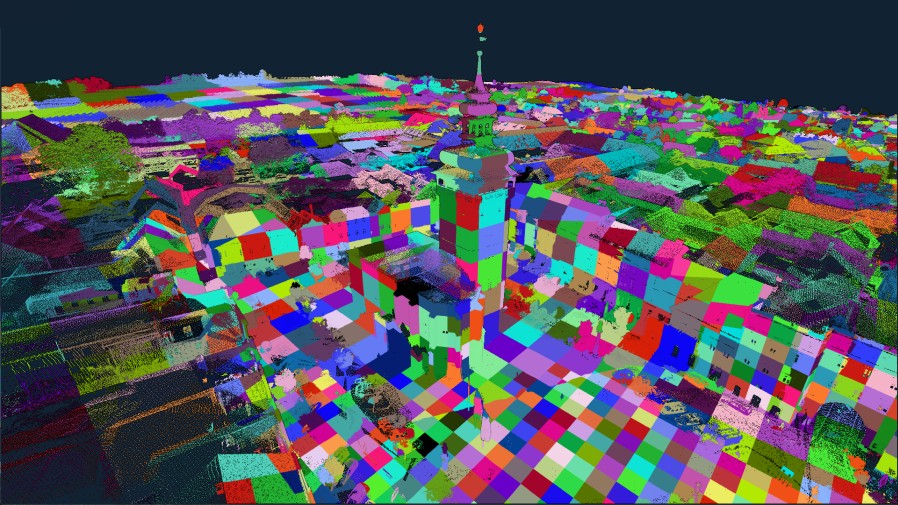}} \vspace*{0.5mm}
  \\ 
  \rotatebox{90}{Morro Bay}  &
     \raisebox{-0.5\totalheight}{\includegraphics[width=0.24\textwidth]{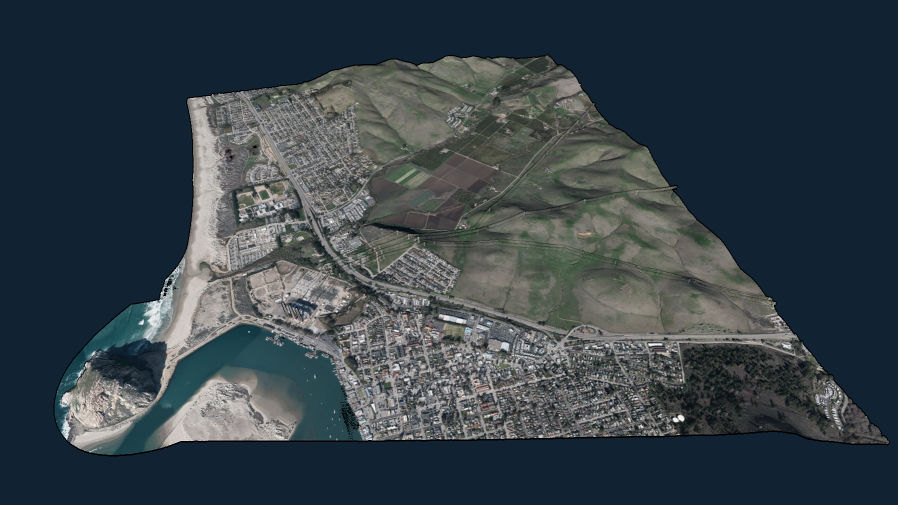}} & 
     \raisebox{-0.5\totalheight}{\includegraphics[width=0.24\textwidth]{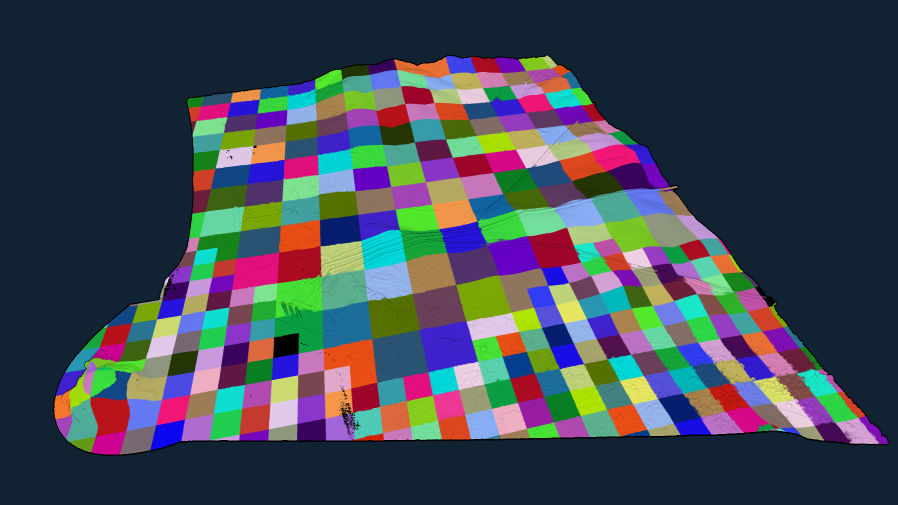}} &
     \raisebox{-0.5\totalheight}{\includegraphics[width=0.24\textwidth]{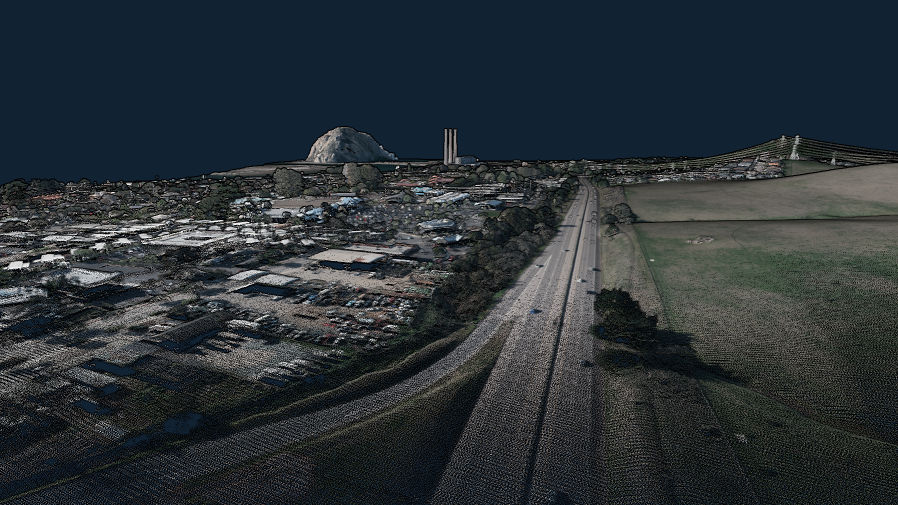}} &
     \raisebox{-0.5\totalheight}{\includegraphics[width=0.24\textwidth]{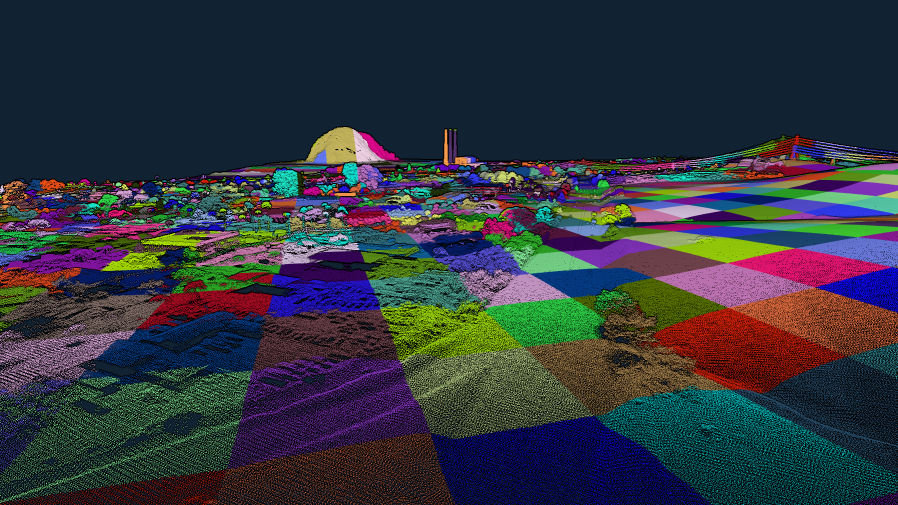}} \vspace*{0.5mm}
  \\ 
  \rotatebox{90}{Meroe}  &
     \raisebox{-0.5\totalheight}{\includegraphics[width=0.24\textwidth]{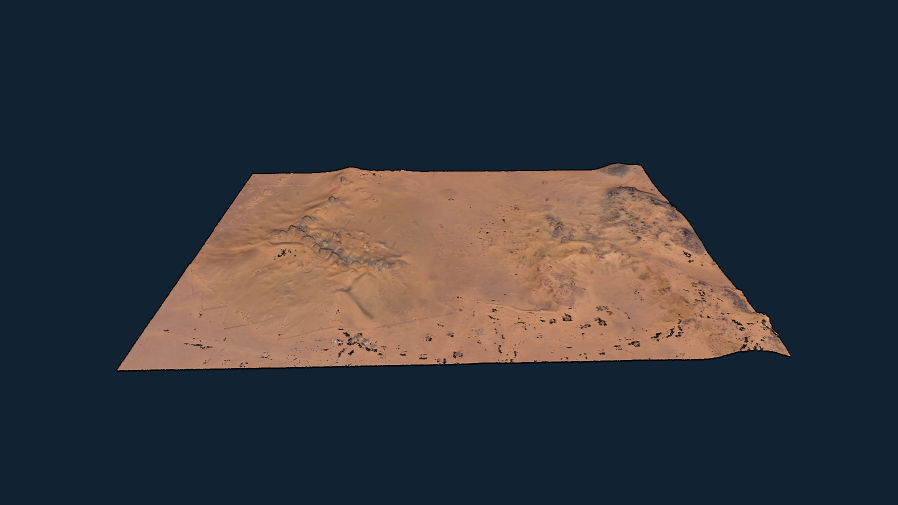}} & 
     \raisebox{-0.5\totalheight}{\includegraphics[width=0.24\textwidth]{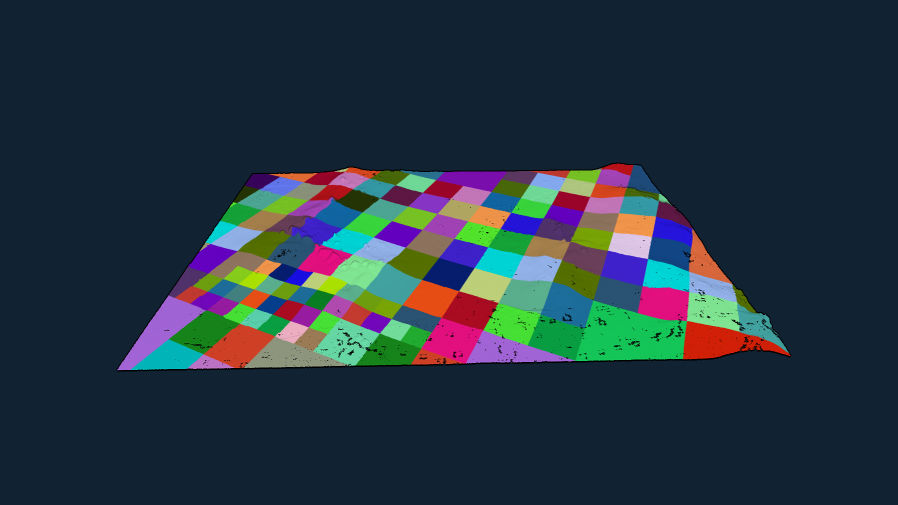}} &
     \raisebox{-0.5\totalheight}{\includegraphics[width=0.24\textwidth]{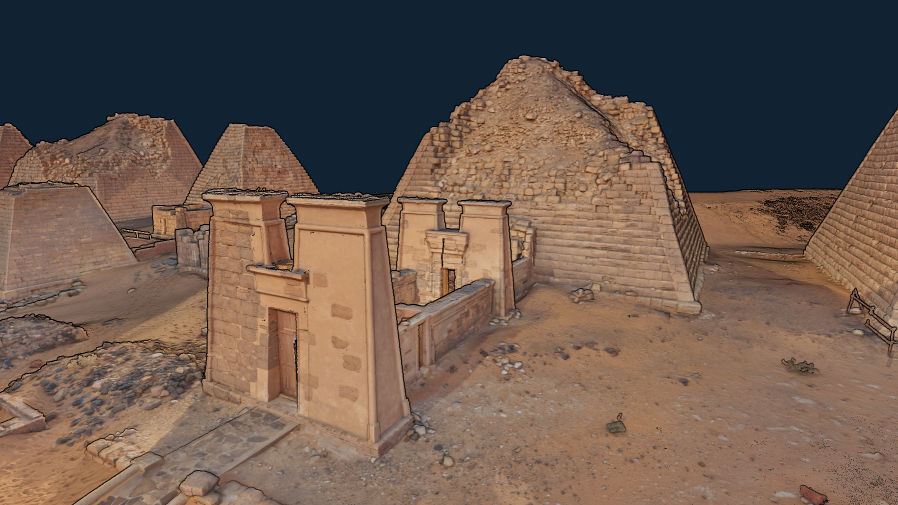}} &
     \raisebox{-0.5\totalheight}{\includegraphics[width=0.24\textwidth]{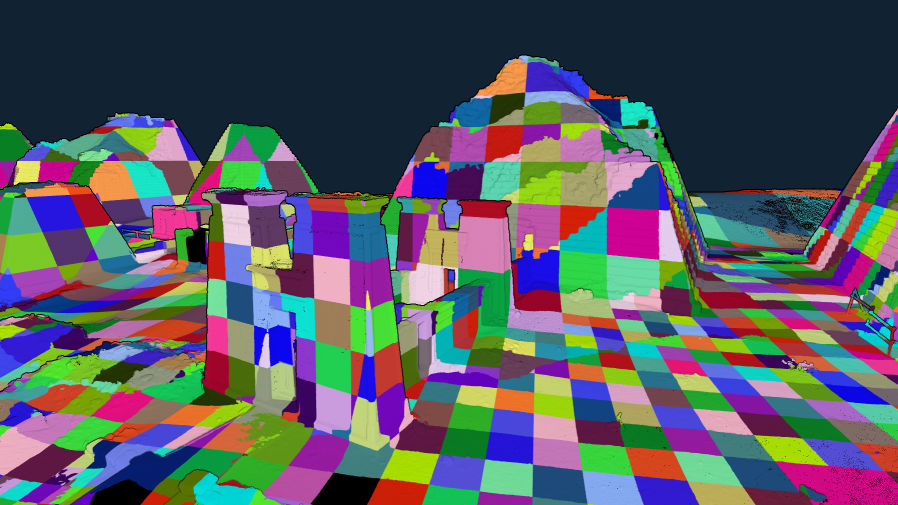}} \vspace*{0.5mm}
  \\ 
  \rotatebox{90}{Endeavor}  &
     \raisebox{-0.5\totalheight}{\includegraphics[width=0.24\textwidth]{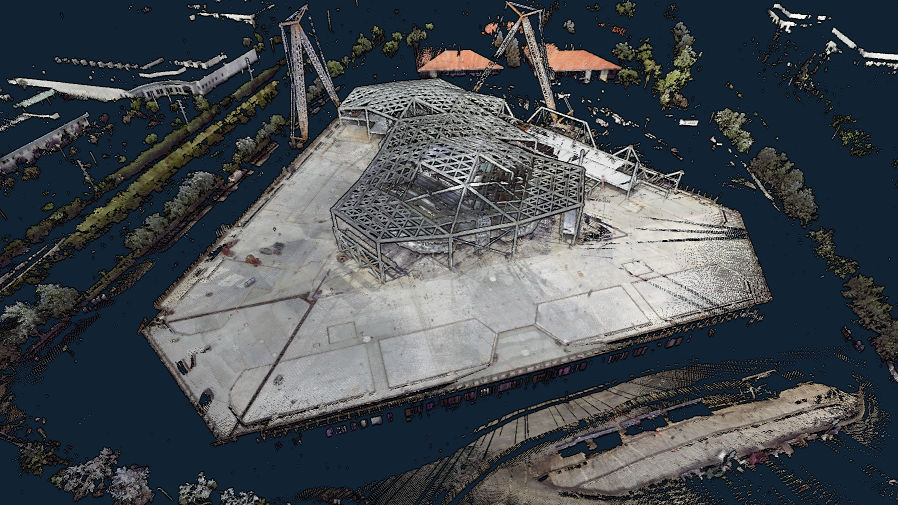}} & 
     \raisebox{-0.5\totalheight}{\includegraphics[width=0.24\textwidth]{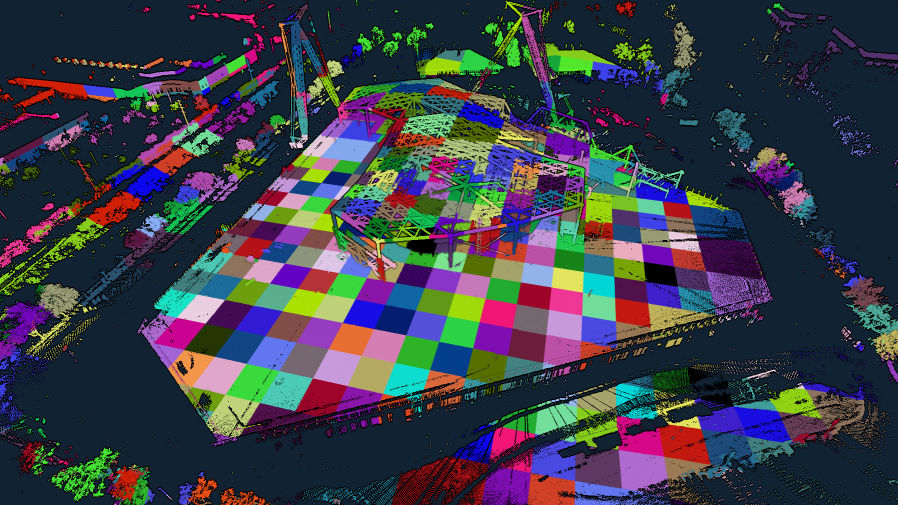}} &
     \raisebox{-0.5\totalheight}{\includegraphics[width=0.24\textwidth]{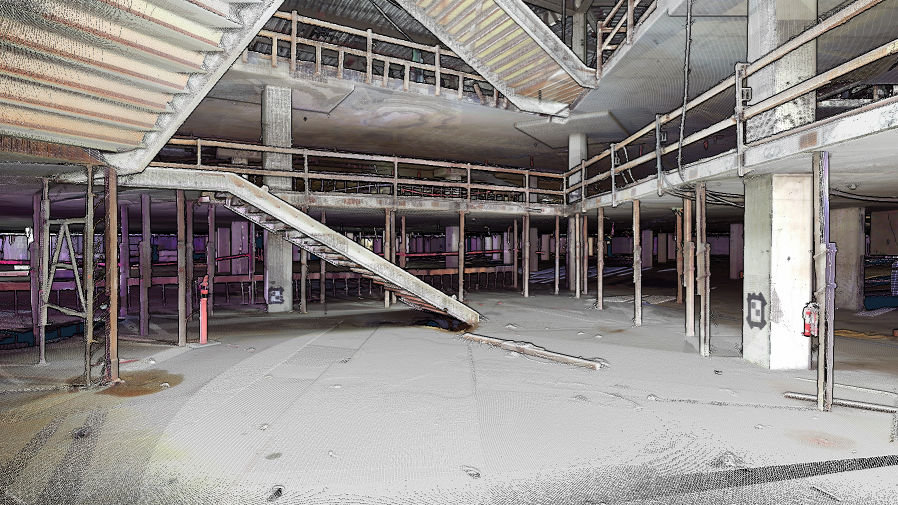}} &
     \raisebox{-0.5\totalheight}{\includegraphics[width=0.24\textwidth]{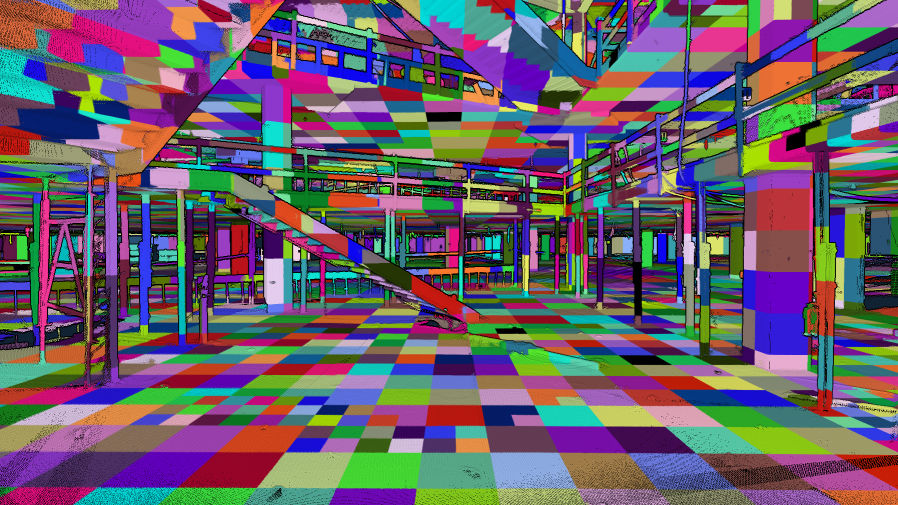}} \\
\end{tabular*}
\caption{Overview and close-ups of our test data sets. The second and fourth columns illustrate the rendered octree nodes.}
\label{fig:test_data_sets}
\end{figure*}
\endgroup

The following systems were used for the evaluation:

\begin{table}[h!]
\begin{tabular*}{\columnwidth}{@{\extracolsep{\fill}} llll}
 \hline
 OS & GPU & CPU & Disk \\ 
 \hline
 Windows 10 & RTX 3060 & Ryzen 7 2700X & Samsung 980 PRO \\ 
 Windows 11 & RTX 4090 & Ryzen 9 7950X & Crucial T700 \\ 
 \hline
\end{tabular*}
\end{table}

Special care was taken to ensure meaningful results for disk IO in our benchmarks: 
\begin{itemize}
    \item On Microsoft Windows, traditional C++ file IO operations such as \texttt{fread} or \texttt{ifstream} are automatically buffered by the operating system. This leads to two issues -- First, it makes the initial access to a file slower and significantly increases CPU usage, which decreases the overall performance of the application and caused stutters when streaming a file from SSD to GPU for the first time. Second, it makes future accesses to the same file faster because the OS now serves it from RAM instead of reading from disk. 
    \item Since we are mostly interested in first-read performance, we implemented file access on Windows via the Windows API's \texttt{ReadFileEx} function together with the \texttt{FILE\_FLAG\_NO\_BUFFERING} flag. It ensures that data is read from disk and also avoids caching it in the first place. As an added benefit, it also reduces CPU usage and resulting stutters. 
\end{itemize}

We evaluated the following performance aspects, with respect to our goal of simultaneously updating the LOD structure and rendering the intermediate results:

\begin{enumerate}
    \item Throughput of the incremental LOD construction in isolation.
    \item Throughput of the incremental LOD construction while streaming points from disk and simultaneously rendering the intermediate results in real time.
    \item Average and maximum duration of all incremental updates.
    \item Performance of rendering nodes up to a certain level of detail.
\end{enumerate}

\subsection{Data Sets}

We evaluated a total of five data sets shown in Figure~\ref{fig:test_data_sets}, three file formats, and Morton-ordering vs. the original ordering by scan position. Chiller and Meroe are photogrammetry-based data sets, Morro Bay is captured via aerial LIDAR, Endeavor via terrestrial laser scans, and Retz via a combination of terrestrial (town center, high-density) and aerial LIDAR (surrounding, low-density). 

The LAS and LAZ file formats are industry-standard point cloud formats. Both store XYZ, RGB, and several other attributes. Due to this, LAS requires either 26 or 34 bytes per point for our data sets. LAZ provides a good and lossless compression down to around 2-5 bytes/point, which is why most massive LIDAR data sets are distributed in that format. However, it is quite CPU-intensive and therefore slow to decode. SIM is a custom file format that stores points in the update kernel's expected format -- XYZRGBA (3 x float + 4 x uint8, 16 bytes per point).

Endeavor is originally ordered by scan position and the timestamp of the points, but we also created a Morton-ordered variation to evaluate the impact of the order. 

\subsection{Construction Performance}

Table~\ref{tab:construction_performance} covers items 1-3 and shows the construction performance of our method on the test systems. The incremental LOD construction kernel itself achieves throughputs of up to 300M points per second on an RTX 3060, and up to 1.2 billion points per second on an RTX 4090. The whole system, including times to stream points from disk and render intermediate results, achieves up to 100 million points per second on an RTX 3060 and up to 580 million points per second on the RTX 4090. The durations of the incremental updates are indicators for the overall impact on fps (average) and occasional stutters (maximum). We implemented a time budget of 10ms per frame to reduce the maximum durations of the update kernel (RTX 3060: 45ms → 16ms; RTX 4090 25ms → 13ms). After the budget is exceeded, the kernel stops and resumes the next frame. Our method benefits from locality as shown by the Morton-ordered variant of the Endeavor data set, which increases the construction performance by a factor of x2.5 (497 MP/s → 1221 MP/s).

\begin{table*}
\begin{tabular*}{\textwidth}{@{\extracolsep{\fill}}|l|c|rrr|rr|rr|rrr|}
\hline 
                       &            & \multicolumn{3}{c|}{}                &  \multicolumn{2}{c|}{Update}   &  \multicolumn{2}{c|}{Duration}  & \multicolumn{3}{c|}{Throughput}               \\
Data Set               &  GPU       & points   & format  & size        &           avg  &      max      &     updates    & total          &     updates   &      total      &   total     \\
                       &            &     (M)  &         &      (GB)   &          (ms)  &      (ms)     &     (sec)      & (sec)          &       (MP/s)  &      (MP/s)     &   (GB/s)    \\
\hline                                                              
Chiller                &  RTX 3060  &  73.6        &     LAS &   1.9       &           1.2  & 13.3          &          0.2   &        1.3     &         297   &           54    &    1.4   \\
                       &            &              &     SIM &   1.2       &           2.0  & 14.0          &          0.2   &        0.8     &         298   &           87    &    1.4   \\
Retz                   &            & 145.5        &     LAS &   4.9       &           1.0  & 14.9          &          0.6   &        3.2     &         260   &           45    &    1.5   \\
                       &            &              &     SIM &   2.3       &           2.8  & 14.3          &          0.5   &        1.6     &         272   &           91    &    1.4   \\
Morro Bay              &            & 350.0        &     LAS &  11.9       &           1.2  & 15.9          &          1.4   &        7.6     &         242   &           46    &    1.6   \\
                       &            &              &     SIM &   5.6       &           4.1  & 16.1          &          1.4   &        3.5     &         247   &          100    &    1.6   \\
\hline
Chiller                &  RTX 4090  &  73.6        &     LAS &   1.9       &           0.6  &  7.5          &          0.1   &        0.3     &       1,215   &          291    &    7.5  \\
                       &            &              &     SIM &   1.2       &           0.6  &  8.0          &          0.1   &        0.2     &       1,217   &          439    &    7.2  \\
Retz                   &            & 145.5        &     LAS &   4.9       &           0.4  &  8.0          &          0.1   &        0.7     &       1,145   &          221    &    7.4  \\
                       &            &              &     SIM &   2.3       &           1.0  &  8.6          &          0.1   &        0.4     &       1,187   &          425    &    6.7  \\
Morro Bay              &            & 350.0        &     LAS &  11.9       &           0.6  &  9.2          &          0.4   &        1.5     &         979   &          234    &    8.0  \\
                       &            &              &     SIM &   5.6       &           1.5  & 10.9          &          0.3   &        0.8     &       1,030   &          458    &    7.3  \\
Meroe                  &            & 684.4        &     LAS &  23.3       &           0.7  & 10.4          &          0.8   &        2.8     &         882   &          241    &    8.2  \\
                       &            &              &     SIM &  11.4       &           1.9  & 12.1          &          0.7   &        1.7     &         945   &          401    &    6.4  \\
Endeavor               &            & 796.0        &     LAS &  20.7       &           7.0  & 12.6          &          1.6   &        2.6     &         497   &          307    &    8.0  \\
                       &            &              &     LAZ &   8.0       &           0.2  &  7.2          &          2.4   &       25.1     &         328   &           32    &    0.3  \\
                       &            &              &     SIM &  12.7       &           9.1  & 12.9          &          1.6   &        2.3     &         497   &          341    &    5.4  \\
Endeavor (z-order)     &            &              &     SIM &  12.7       &           2.2  & 10.7          &          0.7   &        1.4     &       1,221   &          581    &    9.3  \\
\hline
\end{tabular*}
\caption{
LOD Construction Performance showing average and maximum durations of the update kernel, total duration of all updates or the whole system, and the throughput in million points per second (MP/s) or gigabytes per second (GB/s). Total duration includes the time to load points from disk, stream them to the GPU, and insert them into the octree. Update measures the duration of all incremental per-frame (may process multiple batches) updates in isolation. Throughput in GB/s refers to the file size, which depends on the number of points and the storage format (ranging from 10 (LAZ), 16(SIM) to 26 or 34(LAS) byte per point). }
\label{tab:construction_performance}
\end{table*}

\subsection{Rendering Performance}

Regarding rendering performance, we show that linked lists of chunks of points/voxels are suitable for high-performance real-time rendering by rendering the constructed LOD structure at high resolutions (pixel-sized voxels), whereas performance-sensitive rendering engines (targeting browsers, VR, ...) would limit the number of points/voxels of the same structure to a point budget in the single-digit millions, and then fill resulting gaps by increasing point sizes accordingly. Table~\ref{tab:rendering_performance} shows that we are able to render up to 89.4 million pixel-sized points and voxels in 2.7 milliseconds, which leaves the majority of a frame's time for the construction kernel (or higher-quality shading). Table~\ref{tab:chunksizes} shows that the size of chunks has negligible impact on rendering performance (provided they are larger than the workgroup size).

We implemented the atomic-based point rasterization by Schütz et al.~\cite{SCHUETZ-2021-PCC}, including the early-depth test. Compared to their brute-force approach that renders all points in each frame, our on-the-fly LOD approach reduces rendering times on the RTX 4090 by about 5 to 12 times, e.g. Morro Bay is rendered about 5 to 9 times faster (overview: 7.1ms → 0.8ms; closeup: 6.3ms → 1.1ms) and Endeavor is rendered about 5 to 12 times faster (overview: 13.7ms → 1.1ms; closeup: 13.8ms → 2.7ms). Furthermore, the generated LOD structures would allow improving rendering performance even further by lowering the detail to less than 1 point per pixel. In terms of throughput (rendered points/voxels per second), our method is several times slower (Morro Bay overview: 50MP/s → 15.8MP/s; Endeavor overview: 58MP/s → 15.5MP/s). This is likely because throughput dramatically rises with overdraw, i.e., if thousands of points project to the same pixel, they share state and collaboratively update the pixel. 

At this time, we did not implement the approach presented in Sch\"utz et al.'s follow-up paper~\cite{SCHUETZ-2022-PCC} that further improves rendering performance by compressing points and reducing memory bandwidth.

\begin{table*}
\begin{tabular*}{\textwidth}{@{\extracolsep{\fill}}|l|rrrr|rrrr|}
\hline 
                   & \multicolumn{4}{c|}{overview}       & \multicolumn{4}{c|}{closeup}        \\
                   & points & voxels & nodes & duration & points & voxels & nodes & duration \\
\hline 
Chiller            & 1.5 M   & 10.3 M   & 450   & 1.3 ms     & 22.7 M  & 19.5 M  & 1813  & 3.5 ms     \\
Morro Bay          & 0.4 M   & 12.5 M   & 518   & 1.5 ms     &  8.8 M  & 16.2 M  & 1065  & 2.2 ms     \\
Meroe              & 1.4 M   &  2.5 M   & 208   & 0.9 ms     & 24.6 M  & 21.0 M  & 2069  & 3.7 ms     \\
Endeavor           & 5.2 M   & 11.8 M   & 914   & 2.1 ms     & 54.0 M  & 50.7 M  & 4811  & 7.5 ms     \\
\hline 
\end{tabular*}
\caption{Rendering performance from overview and closeup viewpoints shown in Figure~\ref{fig:test_data_sets}. Points and voxels are both rendered as pixel-sized splats. Some voxels may be processed but discarded because they are replaced by higher-res data in visible child nodes. Up to 104 million points+voxels stored in linked-lists inside $4,811$ octree nodes are rasterized at >120fps for close-up views at high levels of detail.}
\label{tab:rendering_performance}
\end{table*}

\begin{table*}
\begin{tabular*}{\textwidth}{@{\extracolsep{\fill}}|l|c|rrrrr|rrrrr|}
\hline 
                   &         & \multicolumn{5}{c|}{overview}       & \multicolumn{5}{c|}{closeup}        \\
                   &   GPU   & points & voxels & nodes & duration & samples/ms & points & voxels & nodes & duration & samples/ms \\
\hline 
Chiller            &  3060   & 2.6 M   &  8.4 M   & 441   & 3.3 ms &  3.3 M    & 28.0 M  &  7.9 M  & 1678  & 7.4 ms & 4.9 M     \\
Retz               &         & 5.2 M   & 12.4 M   & 644   & 4.4 ms &  4.0 M    & 18.9 M  &  7.5 M  & 1616  & 6.0 ms & 4.4 M     \\
Morro Bay          &         & 0.6 M   & 12.0 M   & 477   & 3.5 ms &  3.6 M    & 16.3 M  & 13.7 M  & 1346  & 6.5 ms & 4.6 M     \\
\hline 
Chiller            &  4090   & 2.6 M   &  8.4 M   & 441   & 0.7 ms & 15.7 M    & 28.0 M  &  7.9 M  & 1678  & 1.3 ms & 27.6 M     \\
Retz               &         & 5.2 M   & 12.4 M   & 644   & 0.8 ms & 22.0 M    & 18.9 M  &  7.5 M  & 1616  & 1.1 ms & 24.0 M     \\
Morro Bay          &         & 0.6 M   & 12.0 M   & 477   & 0.8 ms & 15.8 M    & 16.3 M  & 13.7 M  & 1346  & 1.1 ms & 27.3 M     \\
Meroe              &         & 1.9 M   &  2.0 M   & 190   & 0.5 ms &  7.8 M    & 36.4 M  & 17.5 M  & 2500  & 1.9 ms & 28.4 M     \\
Endeavor           &         & 6.5 M   & 10.5 M   & 906   & 1.1 ms & 15.5 M    & 72.7 M  & 16.7 M  & 4956  & 2.7 ms & 33.1 M     \\
\hline 
\end{tabular*}
\caption{Rendering performance from overview and closeup viewpoints shown in Figure~\ref{fig:test_data_sets}. Samples (Points+Voxels) are both rendered as pixel-sized splats.}
\label{tab:rendering_performance}
\end{table*}

\subsection{Chunk Sizes}
\label{sec:chunkSizes}

Table~\ref{tab:chunksizes} shows the impact of chunk sizes on LOD construction and rendering performance. Smaller chunk sizes reduce memory usage but also increase construction duration. The rendering duration, on the other hand, is largely unaffected by the range of tested chunk sizes. We opted for a chunk size of 1k for this paper because it makes our largest data set -- \emph{Endeavor} -- fit on the GPU, and because the slightly better construction kernel performance of larger chunks did not significantly improve the total throughput of the system.

\begin{table}[h!]
\begin{tabular*}{\columnwidth}{@{\extracolsep{\fill}} rrrr}
 \hline
 Chunk Size & construct (ms) & Memory (GB) & rendering (ms) \\ 
 \hline
   500 & 933.9 & 17.1 & 1.9 \\ 
  1\,000 & 734.9 & 17.2 & 1.9 \\ 
  2\,000 & 654.5 & 17.6 & 1.9 \\ 
  5\,000 & 618.1 & 18.9 & 1.9 \\ 
 10\,000 & 611.0 & 21.0 & 1.9 \\ 
 \hline
\end{tabular*}
\caption{The Impact of points/voxels per chunk on total construction duration, memory usage for octree data, and rendering times. (Close-up viewpoint of the \emph{Meroe} data set on an RTX 4090)}
\label{tab:chunksizes}
\end{table}

\subsection{Performance Discussion}

Our approach constructs LOD structures up to 320 times faster than the incremental, CPU-based approach of Bormann et al.~\cite{RealTimeIndexingBormann} (1.8MP/s → 580MP/s) (Peak result of 1.8MP/s taken from Bormann et al.) while points are simultaneously rendered on the same GPU, and up to 677 times faster if we only account for the construction times (1.8MP/s → 1222MP/s). On the same 4090 test system, our incremental approach is about 7.6 times slower than the non-incremental, GPU-based construction method of Schütz et al.~\cite{SCHUETZ-2023-LOD} for the same first-come sampling method (Morro Bay: 7500 MP/s → 979 MP/s), but about 18 times (Morro Bay; LAS; with rendering: 13 MP/s → 234 MP/s) to 75 times (Morro Bay; LAS; construction: 13 MP/s → 979 MP/s) faster than their non-incremental, CPU-based method~\cite{SCHUETZ-2020-MPC}. 

In practice, point clouds are often compressed and distributed as LAZ files. LAZ compression is lossless and effective, but slow to decode. Incremental methods such as Bormann et al.~\cite{RealTimeIndexingBormann} and ours are especially interesting for these as they allow users to immediately see results while loading is in progress. Although non-incremental methods such as \cite{SCHUETZ-2023-LOD} feature a significantly higher throughput of several billions of points per second, they are nevertheless bottle-necked by the 32 million points per second (see Table~\ref{tab:construction_performance}, Endeavor) with which we can load and decompress such data sets, while they can only display the results after loading and processing is fully finished.

\section{Conclusion, Discussion and Potential Improvements}
\label{sec:conclusion}

In this paper, we have shown that GPU-based computing allows us to incrementally construct an LOD structure for point clouds at the rate at which points can be loaded from an SSD, and immediately display the results to the user in real time. Thus, users are able to quickly inspect large data sets right away. without the need to wait until LOD construction is finished. 

There are, however, several limitations and potential improvements that we would like to mention:

\begin{itemize}
    \item \textbf{Out-of-Core}: This approach is currently in-core only and serves as a baseline and a step towards future out-of-core approaches. For arbitrarily large data sets, out-of-core approaches are necessary that flush least-recently-modified-and-viewed nodes to disk. Once they are needed again, they will have to be reloaded -- either because the node becomes visible after camera movement, or because newly loaded points are inserted into previously flushed nodes.
    \item \textbf{Compression}: In-between ``keeping the node's growable data structure in memory" and ``flushing the entire node to disk" is the potential to keep nodes in memory, but convert them into a more efficient structure. ``Least recently modified" nodes can be converted into a non-growable, compressed form with higher memory and rendering efficiency, and ``least recently modified and viewed" nodes could be compressed even further and decoded on-demand for rendering. For reference, voxel coordinates could be encoded relative to voxels in parent nodes, which requires about 2 bit per voxel, and color values of z-ordered voxels could be encoded with BC texture compression~\cite{BC7Format}, which requires about 8 bit per color, for a total of 10 bit per voxel. Currently, our implementation uses 16 bytes (128 bit) per voxel.
    \item \textbf{Color-Filtering}: Our implementation currently does a first-come color sampling for voxels, which leads to aliasing artifacts similar to textured meshes without mipmapping, or in some cases even bias towards the first scan in a collection of multiple overlapping scans. The implementation offers a rendering mode that blends overlapping points~\cite{Botsch:HQS, SCHUETZ-2021-PCC}, which significantly improves quality, but a sparse amount of overlapping points at low LODs are not sufficient to reconstruct a perfect representation of all the missing points from higher LODs. Thus, proper color filtering approaches will need to be implemented to create representative averages at lower levels of detail. We implemented averaging in post-processing~\cite{SCHUETZ-2023-LOD}, but in the future, we would like to integrate color sampling directly into the incremental LOD generation pipeline. The main reason we did not do this yet is the large amount of extra memory that is required to accumulate colors for each voxel, rather than the single bit that is required to select the first color. We expect that this approach can work with the help of least-recently-used queues that help us predict which nodes still need high-quality sampling grids, and which nodes do not need them anymore. This can also be combined with the hash map approach by Wand et al.~\cite{Wand2008}, which reduces the amount of memory of each individual sampling grid. 
    \item \textbf{Quality}: To improve quality, future work in fast and incremental LOD construction may benefit from fitting higher quality point primitives (Surfels, Gaussian Splats, ... \cite{Surfels, SurfaceSplatting, SurfaceSplattingHardware, kerbl3Dgaussians}) to represent lower levels of detail. Considering the throughput of SSDs (up to 580M Points/sec), efficient heuristics to quickly generate and update splats are required, and load balancing schemes that progressively refine the splats closer to the user's current viewpoint. 
    \item \textbf{Sparse Buffers}: An alternative to the linked-list approach for growable arrays of points may be the use of virtual memory management (VMM)~\cite{VirtualMM}. VMM allows allocating large amounts of virtual memory, and only allocates actual physical memory as needed (similar to OpenGL's \texttt{ARB\_sparse\_buffer} extension~\cite{GLSparseBuffer}). Thus, each node could allocate a massive virtual capacity for its points in advance, progressively back it with physical memory as the amount of points we add grows, and thereby make linked lists obsolete. We did not explore this option at this time because our entire update kernel -- including allocation of new nodes, insertion of points and required allocations of additional memory, etc. -- runs on the device, while VMM operations must be called from the host.
\end{itemize}

The source code for this paper is available at \url{https://github.com/m-schuetz/SimLOD}. The repository also contains several subsets of the Morro Bay data set (which in turn is a subset of San Simeon~\cite{CA13}) in different file formats.

\section{Acknowledgements}

The authors wish to thank \emph{Weiss AG} for the \emph{Chiller} data set; \emph{Riegl Laser Measurement Systems} for providing the data set of the town of \emph{Retz}; \emph{PG\&E} and \emph{Open Topography} for the \emph{Morro Bay} data set (a subset of the San Simeon data set)~\cite{CA13}; \emph{Iconem} for the \emph{Northern necropolis - Meroë} data set~\cite{Meroe}; \emph{NVIDIA} for the \emph{Endeavor} data set; \emph{Keenan Crane} for the \emph{Spot} model; and Bunds et al.~\cite{SanAndrasFaultDataset} for the San Andras Fault data set.

This research has been funded by WWTF project \emph{ICT22-055 - Instant Visualization and Interaction for Large Point Clouds}, and
FFG project \emph{LargeClouds2BIM}. 


\printbibliography                

@inproceedings{Surfels,
    author = {Pfister, Hanspeter and Zwicker, Matthias and van Baar, Jeroen and Gross, Markus},
    title = {Surfels: Surface Elements As Rendering Primitives},
    booktitle = {Proceedings of the 27th Annual Conference on Computer Graphics and Interactive Techniques},
    series = {SIGGRAPH '00},
    year = {2000},
    pages = {335--342},
    numpages = {8},

    acmid = {344936},
    publisher = {ACM Press/Addison-Wesley Publishing Co.},
    address = {New York, NY, USA},
}

@inproceedings{QSplat,
    author = {Rusinkiewicz, Szymon and Levoy, Marc},
    title = {QSplat: A Multiresolution Point Rendering System for Large Meshes},
    year = {2000},
    publisher = {ACM Press/Addison-Wesley Publishing Co.},
    address = {USA},

    booktitle = {Proceedings of the 27th Annual Conference on Computer Graphics and Interactive Techniques},
    pages = {343–352},
    numpages = {10},
    series = {SIGGRAPH ’00}
}

@INPROCEEDINGS{Botsch:HQS,
    author={Mario Botsch and Alexander  Hornung and Matthias Zwicker and Leif Kobbelt},
    booktitle={Proceedings Eurographics/IEEE VGTC Symposium Point-Based Graphics, 2005.}, 
    title={High-quality surface splatting on today's GPUs}, 
    year={2005},
    volume={},
    number={},
    pages={17-141},
}

@article{Gnther2013AGP,
    title={A GPGPU-based Pipeline for Accelerated Rendering of Point Clouds},
    author={Christian G{\"u}nther and Thomas Kanzok and Lars Linsen and Paul Rosenthal},
    journal={J. WSCG},
    year={2013},
    volume={21},
    pages={153-161}
}

@inproceedings{MartinezRubi2015,
      title={Taming the beast: Free and open-source massive point cloud web visualization},
      author={Oscar Martinez-Rubi and Stefan Verhoeven and M. van Meersbergen and Markus Schütz and Peter van Oosterom and Romulo Goncalves and T. P. M. Tijssen},
      note = {Capturing Reality Forum 2015, Salzburg, Austria},
      year={2015},

}

@article{GOBBETTI2004,
    author = {Gobbetti, Enrico and Marton, Fabio},
    title = {Layered Point Clouds: A Simple and Efficient Multiresolution Structure for Distributing and Rendering Gigantic Point-sampled Models},
    journal = {Comput. Graph.},
    issue_date = {December, 2004},
    volume = {28},
    number = {6},
    year = {2004},
    pages = {815--826},
    numpages = {12},
    acmid = {1652364},
    publisher = {Pergamon Press, Inc.},
    address = {Elmsford, NY, USA},
}

@misc{AHN2,
    title = {AHN2},
    key = {AHN2},
    note = {\url{https://www.pdok.nl/introductie/-/article/actueel-hoogtebestand-nederland-ahn2-}, Accessed 2021.03.27},
    shorthand = {AHN2},
    author = {},
    year = {}
}

@misc{AHN,
    title = {AHN},
    key = {AHN},
    note = {\url{https://www.ahn.nl/kwaliteitsbeschrijving}, Accessed 2023.06.01},
    shorthand = {AHN},
    author = {},
    year = {}
}

@misc{Entwine,
    title = {Entwine},
    key = {Entwine},
    note = {\url{https://entwine.io/}, Accessed 2021.04.13},
    shorthand = {ENT},
}

@article{Dachsbacher2003,
    author = {Dachsbacher, Carsten and Vogelgsang, Christian and Stamminger, Marc},
    title = {Sequential Point Trees},
    year = {2003},
    issue_date = {July 2003},
    publisher = {Association for Computing Machinery},
    address = {New York, NY, USA},
    volume = {22},
    number = {3},

    journal = {ACM Trans. Graph.},
    pages = {657–662},
    numpages = {6}
}

@article{scheiblauer2011,
    title =      "Out-of-Core Selection and Editing of Huge Point Clouds",
    author =     "Claus Scheiblauer and Michael Wimmer",
    year =       "2011",
    journal =    "Computers \& Graphics",
    number =     "2",
    volume =     "35",
    pages =      "342--351",
}

@article{Wand2008,
	title = "Processing and interactive editing of huge point clouds from 3D scanners",
	journal = "Computers \& Graphics",
	volume = "32",
	number = "2",
	pages = "204 - 220",
	year = "2008",

	author = "Michael Wand and Alexander Berner and Martin Bokeloh and Philipp Jenke and Arno Fleck and Mark Hoffmann and Benjamin Maier and Dirk Staneker and Andreas Schilling and Hans-Peter Seidel",
}

@INPROCEEDINGS{Goswami2010,
    author={P. {Goswami} and Y. {Zhang} and R. {Pajarola} and E. {Gobbetti}},
    booktitle={2010 18th Pacific Conference on Computer Graphics and Applications}, 
    title={High Quality Interactive Rendering of Massive Point Models Using Multi-way kd-Trees}, 
    year={2010},
    volume={},
    number={},

    pages={93-100}
}

@INPROCEEDINGS{Kang2019,
    author="Lai Kang and Jie Jiang and Yingmei Wei and Yuxiang Xie",
    booktitle={2019 IEEE Fourth International Conference on Data Science in Cyberspace (DSC)}, 
    title={Efficient Randomized Hierarchy Construction for Interactive Visualization of Large Scale Point Clouds}, 
    year={2019},
    volume={},
    number={},
    pages={593-597},
}

@inproceedings {Bormann:PCI,
    booktitle = {Smart Tools and Apps for Graphics - Eurographics Italian Chapter Conference},
    editor = {Biasotti, Silvia and Pintus, Ruggero and Berretti, Stefano},
    title = {{A System for Fast and Scalable Point Cloud Indexing Using Task Parallelism}},
    author = {Bormann, Pascal and Krämer, Michel},
    year = {2020},
    publisher = {The Eurographics Association},

}

@article{SCHUETZ-2021-PCC,
	title =      "Rendering Point Clouds with Compute Shaders and Vertex Order
			   Optimization",
	author =     "Markus Schütz and Bernhard Kerbl and Michael Wimmer",
	year =       "2021",
	issn =       "1467-8659",
	journal =    "Computer Graphics Forum",
	number =     "4",
	volume =     "40",
	booktitle =  "techreport",
	pages =      "115--126",
	URL =        "https://www.cg.tuwien.ac.at/research/publications/2021/SCHUETZ-2021-PCC/"
}

@article{SCHUETZ-2022-PCC,
    title =      "Software Rasterization of 2 Billion Points in Real Time",
    author =     "Markus Schütz and Bernhard Kerbl and Michael Wimmer",
    year =       "2022",
    month =      jul,
    journal =    "Proceedings of the ACM on Computer Graphics and Interactive
               Techniques",
    volume =     "5",
    number =     "3",
    pages =      "1--17",
}

@misc{CA13,
  title = {PG\&E Diablo Canyon Power Plant (DCPP): San Simeon and Cambria Faults, CA, Airborne Lidar survey},
  year = {2013},
  author = {{Pacific Gas \& Electric Company}},
  note = {Distributed by OpenTopography},
  shorthand = {CA13}
}

@misc{SanAndrasFaultDataset,
  title = {High Resolution Topography of the Central San Andreas Fault at Dry Lake Valley},
  year = {2020},
  author = {Bunds, M.P. and Scott, C. and Toké, N.A. and Saldivar, J. and Woolstenhulme, L. and Phillips, J. and Keck, M. and Smith, S. and Ranney, M.},
  note = {Distributed by OpenTopography, Accessed 2023.09.29 },
  shorthand = {CA19}
}

@incollection{2015learning,
	author    = {Evans, Alex},
	title     = {{Learning from failure: A Survey of Promising, Unconventional and Mostly Abandoned Renderers for ‘Dreams PS4’, a Geometrically Dense, Painterly UGC Game}},
	booktitle = {ACM SIGGRAPH 2015 Courses, Advances in Real-Time Rendering in Games},
	year      = {2015},
	note = {\url{http://media.lolrus.mediamolecule.com/AlexEvans_SIGGRAPH-2015.pdf} [Accessed 7-June-2022]}
}

@inproceedings {InstantPoints,
    booktitle = {Symposium on Point-Based Graphics},
    editor = {Mario Botsch and Baoquan Chen and Mark Pauly and Matthias Zwicker},
    title = {{Instant Points: Fast Rendering of Unprocessed Point Clouds}},
    author = {Wimmer, Michael and Scheiblauer, Claus},
    year = {2006},
    publisher = {The Eurographics Association},
    ISSN = {1811-7813},
    ISBN = {3-905673-32-0}
}

@article{SCHUETZ-2020-MPC,
    title =      "Fast Out-of-Core Octree Generation for Massive Point Clouds",
    author =     "Markus Schütz and Stefan Ohrhallinger and Michael Wimmer",
    year =       "2020",
    month =      nov,
    issn =       "1467-8659",
    journal =    "Computer Graphics Forum",
    number =     "7",
    volume =     "39",
    pages =      "1--13",
}

@article{Chajdas2014Scalable,
    author =       "Matthäus G. Chajdas and Matthias Reitinger and Rüdiger Westermann",
    title =        "Scalable rendering for very large meshes",
    journal =      "Journal of WSCG",
    year =         "2014",
    volume =       "22",
    pages =        "77--85"
}

@misc{USGS:3DEP,
    title = {{3D Elevation Program (3DEP)}},
    key = {USGS:3DEP},
    note = "\url{https://www.usgs.gov/core-science-systems/ngp/3dep}, Accessed 2020.09.18",
    shorthand = {3DEP},
    author = {},
    year = {}
}

@misc{USGS:Entwine,
    title = {{USGS / Entwine}},
    key = {USGS:Entwine},
    note = {\url{https://usgs.entwine.io}, Accessed 2020.09.18},
    shorthand = {ENTW},
    author = {},
    year = {}
}

@article{NestedIndexing,
    title = {Nested spatial data structures for optimal indexing of LiDAR data},
    journal = {ISPRS Journal of Photogrammetry and Remote Sensing},
    volume = {195},
    pages = {287-297},
    year = {2023},
    issn = {0924-2716},
    author = {Carlos J. Ogayar-Anguita and Alfonso López-Ruiz and Antonio J. Rueda-Ruiz and Rafael J. Segura-Sánchez}
}

@article{VANOOSTEROM2022119,
    title = {Organizing and visualizing point clouds with continuous levels of detail},
    journal = {ISPRS Journal of Photogrammetry and Remote Sensing},
    volume = {194},
    pages = {119-131},
    year = {2022},
    issn = {0924-2716},
    author = {Peter {van Oosterom} and Simon {van Oosterom} and Haicheng Liu and Rod Thompson and Martijn Meijers and Edward Verbree}
}

@INPROCEEDINGS{9671659,
    author={Kocon, Kevin and Bormann, Pascal},
    booktitle={2021 IEEE International Conference on Big Data (Big Data)}, 
    title={Point cloud indexing using Big Data technologies}, 
    year={2021},
    volume={},
    number={},
    pages={109-119}
}

@article{korf1985depth,
    title={Depth-first iterative-deepening: An optimal admissible tree search},
    author={Korf, Richard E},
    journal={Artificial intelligence},
    volume={27},
    number={1},
    pages={97--109},
    year={1985},
    publisher={Elsevier}
}

@misc{VirtualMM,
	author = {Cory Perry and Nikolay Sakharnykh},
	title = {Introducing Low-Level GPU Virtual Memory Management},
	year = {2020},
	url = {https://developer.nvidia.com/blog/introducing-low-level-gpu-virtual-memory-management/},
	note = {Accessed 2023.06.01}
}

@misc{GLSparseBuffer,
	author = {The Khronos Group Inc.},
	title = {ARB\_sparse\_buffer Extension},
	year = {2014},
	url = {https://registry.khronos.org/OpenGL/extensions/ARB/ARB_sparse_buffer.txt/},
	note = {Accessed 2023.06.01}
}

@inproceedings {RealTimeIndexingBormann,
    booktitle = {Computer Graphics and Visual Computing (CGVC)},
    editor = {Peter Vangorp and Martin J. Turner},
    title = {{Real-time Indexing of Point Cloud Data During LiDAR Capture}},
    author = {Bormann, Pascal and Dorra, Tobias and Stahl, Bastian and Fellner, Dieter W.},
    year = {2022},
    publisher = {The Eurographics Association},
    ISBN = {978-3-03868-188-5},
    DOI = {10.2312/cgvc.20221173}
}

@misc{SCHUETZ-2023-LOD,
    title =      "GPU-Accelerated LOD Generation for Point Clouds",
    author =     "Markus Schütz and Bernhard Kerbl and Philip Klaus and Michael Wimmer",
    year =       "2023",
    month =      feb,
    URL =        "https://www.cg.tuwien.ac.at/research/publications/2023/SCHUETZ-2023-LOD/",
}

@INPROCEEDINGS{PersistentThreads,
    author={Gupta, Kshitij and Stuart, Jeff A. and Owens, John D.},
    booktitle={2012 Innovative Parallel Computing (InPar)}, 
    title={A study of Persistent Threads style GPU programming for GPGPU workloads}, 
    year={2012},
    volume={},
    number={},
    pages={1-14},
    doi={10.1109/InPar.2012.6339596}
}

@misc{CooperativeGroups,
	author = {Mark Harris and Kyrylo Perelygin},
	title = {Cooperative Groups: Flexible CUDA Thread Programming},
	year = {2017},
	url = {https://developer.nvidia.com/blog/cooperative-groups/},
	note = {Accessed 2023.06.05}
}

@article{GpuSoftwarePipeline,
    author = {Kenzel, Michael and Kerbl, Bernhard and Schmalstieg, Dieter and Steinberger, Markus},
    title = {A High-Performance Software Graphics Pipeline Architecture for the GPU},
    year = {2018},
    issue_date = {August 2018},
    publisher = {Association for Computing Machinery},
    address = {New York, NY, USA},
    volume = {37},
    number = {4},
    issn = {0730-0301},
    url = {https://doi.org/10.1145/3197517.3201374},
    doi = {10.1145/3197517.3201374},
    month = {jul},
    articleno = {140},
    numpages = {15}
}

@misc{Meroe,
    title = {Northern necropolis - Meroë},
    key = {Meroe},
    url = {https://app.iconem.com/#/3d/project/public/6384d382-5e58-4454-b8e6-dec45b6e6078/scene/c038fb6f-c16b-421e-a2ac-f7292f1b1c64/},
    note = {Accessed 2023.06.05},
    shorthand = {Meroe},
    author = {Iconem},
    year = {}
}

@inproceedings{ModifyVoxels,
    title={Interactively modifying compressed sparse voxel representations},
    author={Careil, Victor and Billeter, Markus and Eisemann, Elmar},
    booktitle={Computer Graphics Forum},
    volume={39},
    number={2},
    pages={111--119},
    year={2020},
    organization={Wiley Online Library}
}

@misc{BC7Format,
    title = {BC7 Format},
    key = {BC},
    url = {https://learn.microsoft.com/en-us/windows/win32/direct3d11/bc7-format#bc7-implementation/},
    note = {Accessed 2023.06.09},
    shorthand = {BC},
    author = {Microsoft},
    year = {2022}
}

@Article{kerbl3Dgaussians,
      author       = {Kerbl, Bernhard and Kopanas, Georgios and Leimk{\"u}hler, Thomas and Drettakis, George},
      title        = {3D Gaussian Splatting for Real-Time Radiance Field Rendering},
      journal      = {ACM Transactions on Graphics},
      number       = {4},
      volume       = {42},
      month        = {July},
      year         = {2023},
      url          = {https://repo-sam.inria.fr/fungraph/3d-gaussian-splatting/}
}

@inproceedings{SurfaceSplatting,
    author = {Zwicker, Matthias and Pfister, Hanspeter and van Baar, Jeroen and Gross, Markus},
    title = {Surface Splatting},
    year = {2001},
    isbn = {158113374X},
    publisher = {Association for Computing Machinery},
    address = {New York, NY, USA},
    url = {https://doi.org/10.1145/383259.383300},
    booktitle = {Proceedings of the 28th Annual Conference on Computer Graphics and Interactive Techniques},
    pages = {371–378},
    numpages = {8},
    series = {SIGGRAPH '01}
}

@article{SurfaceSplattingHardware,
    author = {Weyrich, Tim and Heinzle, Simon and Aila, Timo and Fasnacht, Daniel B. and Oetiker, Stephan and Botsch, Mario and Flaig, Cyril and Mall, Simon and Rohrer, Kaspar and Felber, Norbert and Kaeslin, Hubert and Gross, Markus},
    title = {A Hardware Architecture for Surface Splatting},
    year = {2007},
    issue_date = {July 2007},
    publisher = {Association for Computing Machinery},
    address = {New York, NY, USA},
    volume = {26},
    number = {3},
    issn = {0730-0301},
    url = {https://doi.org/10.1145/1276377.1276490},
    doi = {10.1145/1276377.1276490},
    journal = {ACM Trans. Graph.},
    month = {jul},
    pages = {90–es},
    numpages = {12}
}

@article{yang_real-time_2010,
    title = {Real-{Time} {Concurrent} {Linked} {List} {Construction} on the {GPU}},
    volume = {29},
    copyright = {© 2010 The Author(s) Journal compilation © 2010 The Eurographics Association and Blackwell Publishing Ltd.},
    issn = {1467-8659},
    url = {https://onlinelibrary.wiley.com/doi/abs/10.1111/j.1467-8659.2010.01725.x},
    language = {en},
    number = {4},
    urldate = {2023-09-19},
    journal = {Computer Graphics Forum},
    author = {Yang, Jason C. and Hensley, Justin and Grün, Holger and Thibieroz, Nicolas},
    year = {2010},
    note = {\_eprint: https://onlinelibrary.wiley.com/doi/pdf/10.1111/j.1467-8659.2010.01725.x},
    pages = {1297--1304},
}

@INPROCEEDINGS{BufferCacheManagement,
    author={Jinhyuk Yoon and Sang Lyul Min and Yookun Cho},
    booktitle={Proceedings International Symposium on Parallel Architectures, Algorithms and Networks. I-SPAN'02}, 
    title={Buffer cache management: predicting the future from the past}, 
    year={2002},
    volume={},
    number={},
    pages={105-110},
    doi={10.1109/ISPAN.2002.1004268}
}

@INPROCEEDINGS{964491,
    author={Cohen, J.D. and Aliaga, D.G. and Weiqiang Zhang},
    booktitle={Proceedings Visualization, 2001. VIS '01.}, 
    title={Hybrid simplification: combining multi-resolution polygon and point rendering}, 
    year={2001},
    volume={},
    number={},
    pages={37-539},
    doi={10.1109/VISUAL.2001.964491}
}

@inproceedings {10.2312:EGWR:EGWR02:043-052,
    booktitle = {Eurographics Workshop on Rendering},
    editor = {P. Debevec and S. Gibson},
    title = {{Hardware-Accelerated Point-Based Rendering of Complex Scenes}},
    author = {Coconu, Liviu and Hege, Hans-Christian},
    year = {2002},
    publisher = {The Eurographics Association},
    ISSN = {1727-3463},
    ISBN = {1-58113-534-3},
    DOI = {10.2312/EGWR/EGWR02/043-052}
}

\end{document}